\documentclass[twocolumn,aps,showpacs]{revtex4-1}

\usepackage{graphicx}
\begin{document}

\title{Nonclassicality and entanglement criteria for bipartite optical fields characterized by quadratic detectors}

\author{Jan Pe\v{r}ina Jr.}
\email{jan.perina.jr@upol.cz}
\author{}
\affiliation{RCPTM, Joint Laboratory of Optics of Palack\'{y}
University and Institute of Physics of the Czech Academy of
Sciences, Faculty of Science, Palack\'{y} University, 17.
listopadu 12, 77146 Olomouc, Czech Republic}
\author{Ievgen I. Arkhipov}
\affiliation{RCPTM, Joint Laboratory of Optics of Palack\'{y}
University and Institute of Physics of the Czech Academy of
Sciences, Faculty of Science, Palack\'{y} University, 17.
listopadu 12, 77146 Olomouc, Czech Republic}
\author{V\'{a}clav Mich\'{a}lek}
\affiliation{Institute of Physics of the Czech Academy of
Sciences, Joint Laboratory of Optics of Palack\'{y} University and
Institute of Physics of CAS, 17. listopadu 50a, 772 07 Olomouc,
Czech Republic}
\author{Ond\v{r}ej Haderka}
\affiliation{Institute of Physics of the Czech Academy of
Sciences, Joint Laboratory of Optics of Palack\'{y} University and
Institute of Physics of CAS, 17. listopadu 50a, 772 07 Olomouc,
Czech Republic}

\begin{abstract}
Numerous inequalities involving moments of integrated intensities
and revealing nonclassicality and entanglement in bipartite
optical fields are derived using the majorization theory,
non-negative polynomials, the matrix approach, as well as the
Cauchy-Schwarz inequality. Different approaches for deriving these
inequalities are compared. Using the experimental photocount
histogram generated by a weak noisy twin beam monitored by a
photon-number-resolving iCCD camera the performance of the derived
inequalities is compared. A basic set of ten inequalities suitable
for monitoring the entanglement of a twin beam is suggested.
Inequalities involving moments of photocounts (photon numbers) as
well as those containing directly the elements of photocount
(photon-number) distributions are also discussed as a tool for
revealing nonclassicality.
\end{abstract}

 \maketitle

\section{Introduction}

The notion of a nonclassical field has been rigorously defined
once the famous Glauber-Sudarshan representation of the density
matrix of an optical field was formulated
\cite{Glauber1963,Sudarshan1963}. From that time, any optical
field with a non-positive Glauber-Sudarshan quasi-distribution is
considered as nonclassical
\cite{Perina1991,Mandel1995,Vogel2001,Dodonov2003}. The analysis
of more complex optical fields involving several optical modes has
shown that one of the reasons for field's nonclassicality is the
presence of quantum correlations (entanglement) among the modes
that constitute the field. As the entanglement is interesting both
for fundamental reasons and various applications (in metrology,
quantum-key distribution, etc.), it has been extensively studied
in the last tens' years in numerous publications. The simplest
case of entanglement between two fields has naturally attracted
the greatest attention. In this case, even the quantification of
entanglement has been found using the Schmidt number for pure
states \cite{Fedorov2014} and its generalization to mixed states
based on finding the closest pure entangled state. Also an
alternative quantification derived from the shape of the Wigner
function has been given \cite{Kenfack2004}. Unfortunately, these
theoretical approaches are difficult to be applied to experimental
optical fields \cite{Chan2007,Just2013}. From the experimental
point of view, joint homodyne tomography
\cite{Leonhardt1997,Lvovsky2009} of both fields is needed to
reveal the joint phase-space quasi-distribution of these fields
and, subsequently, quantify the entanglement via the mentioned
theoretical approaches.

Large experimental demands of entanglement quantification lead to
a simpler concept of entanglement witnesses (criteria) when
dealing with the entanglement. An entanglement witness is a
physical quantity which identifies the entanglement qualitatively
through its values. Typically, such quantity is constructed from
an inequality fulfilled by any classical optical field. A
well-known and frequently-used PPT criterion
\cite{Peres1996,Horodecki1996} represents an entanglement witness
that exploits the eigenvalues of a certain matrix. For specific
systems, this witness can even be converted into an entanglement
measure called negativity \cite{Hill1997}. There exists in
principle an infinite number of entanglement witnesses. On the
other hand, some of these witnesses are more important (or useful)
for the physical reasons. These reasons are pragmatic and they are
related to their performance in the experimental characterization
of optical fields. As quadratic optical detectors are by-far the
most frequently used detectors in optical laboratories worldwide,
the witnesses exploiting the moments of integrated intensity
(farther only intensity) are extraordinarily important
\cite{Haderka2005a,Allevi2012a,Allevi2013,Sperling2015,Harder2016}.
We note that the measurement of the whole joint photocount
distribution of a bipartite optical field can be used to
reconstruct the joint quasi-distribution of integrated intensities
\cite{Perina1991,Haderka2005a,PerinaJr2013a} and to reveal its
negative values observed for nonclassical states.

Here, we theoretically as well as experimentally analyze the
witnesses that indicate negative values of the Glauber-Sudarshan
quasi-distribution of intensities. When applied to the whole
optical field they represent global nonclassicality criteria
(GNCCa). On the other hand, they serve as local nonclassicality
criteria (LNCCa) in the cases of marginal fields describing
individual optical modes. For bipartite optical fields with
classical constituents, the GNCCa represent also entanglement
witnesses (criteria). The reason is that the global
nonclassicality in general reflects either local nonclassicalities
of the constituents or entanglement between the constituents or
both. Twin beams with their signal and idler beams containing many
photon pairs represent a typical example of such bipartite optical
fields. The GNCCa and LNCCa are derived by several approaches that
use the majorization theory \cite{Marshall2010}, consider
non-negative polynomials and quadratic forms (the matrix approach)
\cite{Vogel2008,Miranowicz2010} and exploit the Cauchy-Schwarz
inequality. Relying on the Mandel photodetection formula
\cite{Perina1991,Mandel1995} the corresponding inequalities among
the elements of the joint photocount and photon-number
distributions are also revealed. The performance of the derived
GNCCa is tested on the experimental data characterizing a twin
beam with around nine photon pairs on average and acquired by an
intensified CCD (iCCD) camera. In this case, the GNCCa are also
entanglement criteria.

The paper is organized as follows. In Sec.~II, we give the
simplest inequalities among intensity moments. More complex
inequalities including multiple intensity moments are derived in
Sec.~III using different approaches. Inequalities using the
elements of the joint photocount and photon-number distributions
are discussed in Sec.~IV, together with some useful inequalities
containing photocount and photon-number moments. Sec.~V is devoted
to the application of the derived inequalities to an experimental
noisy twin beam. Conclusions are drawn in Sec.~VI. Additional
inequalities for identifying nonclassicality, that are redundant
to those written in the main text, are summarized in Appendix~A
for completeness.

\section{Simple nonclassicality criteria using intensity moments}

We consider a bipartite optical field composed of in general two
entangled fields, that we call the signal and idler fields and
that have intensities $ W_{\rm s} $ and $ W_{\rm i} $,
respectively. The overall field is described by the joint
signal-idler intensity quasi-distribution $ P_{\rm si}(W_{\rm
s},W_{\rm i}) $ \cite{Perina2005} that allows to determine the
normally-ordered (intensity) moments \cite{Perina1991} along the
relation:
\begin{eqnarray}   
 \langle W_{\rm s}^k W_{\rm i}^l \rangle \hspace{-.5mm}&=&\hspace{-.5mm} \int_{0}^{\infty} dW_{\rm s}
  \int_{0}^{\infty} dW_{\rm i}\, W_{\rm s}^k W_{\rm i}^l P_{\rm si}(W_{\rm s},W_{\rm i}), \nonumber \\
 & & \hspace{10mm} k,l=0,1,\ldots .
\label{1}
\end{eqnarray}

According to the majorization theory applied to polynomials
written in two independent variables \cite{Marshall2010,Lee1990b},
these intensity moments fulfill certain classical inequalities.
Their negation gives us the following series of \emph{Global
NonClassicality Criteria:}
\begin{equation}   
 \sum_{\{k,l\}} \langle W_{\rm s}^k W_{\rm i}^l \rangle <
  \sum_{\{k',l'\}} \langle W_{\rm s}^{k'} W_{\rm i}^{l'} \rangle
\label{2}
\end{equation}
where the summation is performed over all possible permutations of
the indices and the indices $ k,l $ majorize the indices $ k',l' $
($ \{k,l\} \succ \{k',l'\} $). We note that such GNCCa are
obtained in a general form of sum (and difference) of mean values.

To understand in detail the structure of such GNCCa, we explicitly
write those containing the intensity moments up to the fifth order
in the form that naturally arises in the majorization theory:
\begin{eqnarray}   
 & \langle W_{\rm s}^2\rangle + \langle W_{\rm i}^2 \rangle < 2 \langle W_{\rm s}
  W_{\rm i}\rangle, & \label{3} \\
 & \langle W_{\rm s}^3\rangle + \langle W_{\rm i}^3 \rangle < \langle W_{\rm s}^2
  W_{\rm i}\rangle + \langle W_{\rm s} W_{\rm i}^2\rangle, & \label{4} \\
 & \langle W_{\rm s}^4\rangle + \langle W_{\rm i}^4 \rangle < \langle W_{\rm s}^3
  W_{\rm i}\rangle + \langle W_{\rm s} W_{\rm i}^3\rangle, & \label{5} \\
 & \langle W_{\rm s}^4\rangle + \langle W_{\rm i}^4 \rangle < 2\langle W_{\rm s}^2
  W_{\rm i}^2\rangle, & \label{6} \\
 & \langle W_{\rm s}^3 W_{\rm i}\rangle + \langle W_{\rm s} W_{\rm i}^3\rangle < 2\langle W_{\rm s}^2
  W_{\rm i}^2\rangle, & \label{7} \\
 & \langle W_{\rm s}^5\rangle + \langle W_{\rm i}^5 \rangle < \langle W_{\rm s}^4
  W_{\rm i}\rangle + \langle W_{\rm s} W_{\rm i}^4\rangle, & \label{8} \\
 & \langle W_{\rm s}^5\rangle + \langle W_{\rm i}^5 \rangle < \langle W_{\rm s}^3
  W_{\rm i}^2\rangle + \langle W_{\rm s}^2 W_{\rm i}^3\rangle, & \label{9} \\
 & \langle W_{\rm s}^4 W_{\rm i}\rangle + \langle W_{\rm s} W_{\rm i}^4\rangle < \langle W_{\rm s}^3
  W_{\rm i}^2\rangle + \langle W_{\rm s}^2 W_{\rm i}^3\rangle. & \label{10}
\end{eqnarray}
However, the inequalities in Eqs. (\ref{3})---(\ref{10}) can be
recast in turn into the following ones:
\begin{eqnarray}   
 & \langle ( W_{\rm s}- W_{\rm i})^2 \rangle < 0, & \label{11} \\
 & \langle ( W_{\rm s}+W_{\rm i})( W_{\rm s}- W_{\rm i})^2 \rangle < 0, & \label{12} \\
 & \langle ( W_{\rm s}^2 + W_{\rm s}W_{\rm i} + W_{\rm i}^2)( W_{\rm s}- W_{\rm i})^2 \rangle < 0, & \label{13} \\
 & \langle ( W_{\rm s}^2 + 2W_{\rm s}W_{\rm i} + W_{\rm i}^2)( W_{\rm s}- W_{\rm i})^2 \rangle < 0, & \label{14} \\
 & \langle W_{\rm s}W_{\rm i}( W_{\rm s}- W_{\rm i})^2 \rangle < 0, & \label{15} \\
 & \langle ( W_{\rm s} + W_{\rm i})( W_{\rm s}^2 + W_{\rm i}^2)( W_{\rm s}- W_{\rm i})^2 \rangle < 0, & \label{16} \\
 & \langle ( W_{\rm s} + W_{\rm i})( W_{\rm s}^2 + W_{\rm s}W_{\rm i} + W_{\rm i}^2)( W_{\rm s}- W_{\rm i})^2 \rangle < 0, &\label{17} \\
 & \langle ( W_{\rm s} + W_{\rm i})W_{\rm s}W_{\rm i}( W_{\rm s}- W_{\rm i})^2 \rangle < 0. & \label{18}
\end{eqnarray}
A common property of these inequalities is that they are symmetric
with respect to the exchange of indices s and i. This has its
origin in the majorization theory.

Natural generalization of the above GNCCa that removes this
symmetry and that is based upon mean values of non-negative
polynomials is written in the form of the following \emph{Global
NonClassicality Criteria:}
\begin{eqnarray}  
 \langle W_{\rm s}^k W_{\rm i}^l (W_{\rm s}-W_{\rm i})^{2m} \rangle
  &<&0, \nonumber \\
 & & \hspace{-15mm} k,l=0,1,\ldots, \;\; m=1,2\ldots.
\label{19}
\end{eqnarray}

Considering $ m=1$ in Eq.~(\ref{19}) and intensity moments up to
the fifth order, we may define the following GNCCa $ E $:
\begin{eqnarray}  
 E_{001} &\equiv& \langle W_{\rm s}^2\rangle + \langle W_{\rm i}^2 \rangle - 2\langle W_{\rm s}
  W_{\rm i}\rangle <0, \label{20} \\
 E_{101} &\equiv& \langle W_{\rm s}^3\rangle + \langle W_{\rm s}W_{\rm i}^2\rangle - 2\langle W_{\rm s}^2
  W_{\rm i}\rangle <0, \label{21} \\
 E_{011} &\equiv& \langle W_{\rm i}^3\rangle + \langle W_{\rm s}^2W_{\rm i}\rangle - 2\langle W_{\rm s}
  W_{\rm i}^2\rangle <0, \label{22} \\
 E_{201} &\equiv& \langle W_{\rm s}^4\rangle + \langle W_{\rm s}^2W_{\rm i}^2\rangle - 2\langle W_{\rm s}^3
  W_{\rm i}\rangle <0, \label{23} \\
 E_{021} &\equiv& \langle W_{\rm i}^4\rangle + \langle W_{\rm s}^2W_{\rm i}^2\rangle - 2\langle W_{\rm s}
  W_{\rm i}^3\rangle <0, \label{24} \\
 E_{111} &\equiv& \langle W_{\rm s}^3W_{\rm i}\rangle + \langle W_{\rm s}W_{\rm i}^3\rangle -
  2\langle W_{\rm s}^2W_{\rm i}^2\rangle <0, \label{25} \\
 E_{301} &\equiv& \langle W_{\rm s}^5\rangle + \langle W_{\rm s}^3W_{\rm i}^2\rangle - 2\langle W_{\rm s}^4
  W_{\rm i}\rangle <0, \label{26} \\
 E_{031} &\equiv& \langle W_{\rm i}^5\rangle + \langle W_{\rm s}^2W_{\rm i}^3\rangle - 2\langle W_{\rm s}
  W_{\rm i}^4\rangle <0, \label{27} \\
 E_{211} &\equiv& \langle W_{\rm s}^4W_{\rm i}\rangle + \langle W_{\rm s}^2W_{\rm i}^3\rangle -
  2\langle W_{\rm s}^3W_{\rm i}^2\rangle <0, \label{28} \\
 E_{121} &\equiv& \langle W_{\rm s}W_{\rm i}^4\rangle + \langle W_{\rm s}^3W_{\rm i}^2\rangle -
  2\langle W_{\rm s}^2W_{\rm i}^3\rangle <0. \label{29}
\end{eqnarray}
The original GNCCa given in Eqs.~(\ref{3})---(\ref{10}) represent
a subset of the GNCCa written in Eqs.~(\ref{20})---(\ref{29}). In
detail, the GNCCa in Eqs.~(\ref{3})---(\ref{10}) are expressed in
turn as $ E_{001} $, $ E_{101} + E_{011} $, $ E_{201} + E_{111} +
E_{021} $, $ E_{201} + 2E_{111} + E_{021} $, $ E_{111} $, $
E_{301} + E_{211} + E_{121} + E_{031} $, $ E_{301} + 2E_{211} +
2E_{121} + E_{031} $, and $ E_{211} + E_{121} $.

Moreover, the consideration of $ m=2 $ in Eq.~(\ref{19}) gives us
additional three GNCCa:
\begin{eqnarray}  
 E_{002} &\equiv& \langle W_{\rm s}^4\rangle - 4\langle W_{\rm s}^3W_{\rm i}\rangle
   + 6\langle W_{\rm s}^2W_{\rm i}^2\rangle - 4\langle W_{\rm s}W_{\rm i}^3\rangle \nonumber \\
  & &  \mbox{} + \langle W_{\rm i}^4 \rangle <0, \label{30} \\
 E_{102} &\equiv& \langle W_{\rm s}^5\rangle - 4\langle W_{\rm s}^4W_{\rm i}\rangle
   + 6\langle W_{\rm s}^3W_{\rm i}^2\rangle - 4\langle W_{\rm s}^2W_{\rm i}^3\rangle \nonumber \\
  & &  \mbox{} + \langle W_{\rm s}W_{\rm i}^4 \rangle <0, \label{31} \\
 E_{012} &\equiv& \langle W_{\rm s}^4W_{\rm i}\rangle - 4\langle W_{\rm s}^3W_{\rm i}^2\rangle
   + 6\langle W_{\rm s}^2W_{\rm i}^3\rangle - 4\langle W_{\rm s}W_{\rm i}^4\rangle \nonumber \\
  & & \mbox{} + \langle W_{\rm i}^5 \rangle <0, \label{32}
\end{eqnarray}
These GNCCa can be expressed as linear combinations of some of the
GNCCa written in Eqs.~(\ref{20})---(\ref{29}):
\begin{eqnarray} 
 E_{002} &=& E_{201} + E_{021} -2E_{111} , \nonumber \\
 E_{102} &=& E_{301} + E_{121} -2E_{211} , \nonumber \\
 E_{012} &=& E_{211} + E_{031} -2E_{121}.
\label{33}
\end{eqnarray}
As negative signs occur in the combinations of GNCCa $ E $ on the
right-hand-sides of Eqs.~(\ref{33}), the GNCCa $ E_{002} $, $
E_{102} $ and $ E_{012} $ are nontrivial and enrich the set of
GNCCa given in Eqs.~(\ref{20})---(\ref{29}). We note that
analogical situation is met for $ m>2 $ in Eq.~(\ref{19}) and
higher-order intensity moments.

\section{Nonclassicality criteria containing multiple intensity moments}

In this section, we derive the non-classicality criteria that
involve products of intensity moments. We concentrate our
attention to the GNCCa containing products of two intensity
moments, though several GNCCa encompassing also products of three
intensity moments are mentioned. To determine such GNCCa we first
apply the majorization theory. Then we exploit non-negative
polynomials to arrive at additional GNCCa. For completeness, we
mention the GNCCa reached by the matrix approach, that uses
non-negative quadratic forms, and those derived from the
Cauchy-Schwarz inequality. In parallel, we also reveal LNCCa
containing intensity moments and provided by the majorization
theory.

\subsection{Nonclassicality criteria based on the majorization
theory}

We use again the formulas of the majorization theory
\cite{Marshall2010}, now in the systematic way. We begin with the
majorization theory applied to polynomials written in two
independent variables $ W_{\rm s} $ and $ W_{\rm i} $. Contrary to
the approach of the previous section, we make averaging with the
factorized quasi-distribution function $ P_{\rm s}(W_{\rm
s})P_{\rm i}(W_{\rm i}) $ where $ P_{\rm s} $ ($ P_{\rm i} $)
stands for the signal (idler) reduced quasi-distribution function.
The original Eq.~(\ref{2}) attains in this case the form of the
following \emph{Local NonClassicality Criteria}:
\begin{equation}   
 \sum_{\{k,l\}} \langle W_{\rm s}^k \rangle \langle W_{\rm i}^l \rangle
  < \sum_{\{k',l'\}} \langle W_{\rm s}^{k'} \rangle \langle W_{\rm i}^{l'}
  \rangle;
\label{34}
\end{equation}
$ \{k,l\} \succ \{k',l'\} $. Considering intensity moments up to
the fifth order, we arrive at the following six LNCCa expressed in
terms of the intensity moments of the local signal and idler
fields:
\begin{eqnarray}  
 B^{20}_{11} &\equiv& \langle W_{\rm s}^2\rangle + \langle W_{\rm i}^2\rangle -
   2\langle W_{\rm s}\rangle \langle W_{\rm i} \rangle < 0 , \label{35}\\
 B^{30}_{21} &\equiv& \langle W_{\rm s}^3\rangle + \langle W_{\rm i}^3\rangle -
   \langle W_{\rm s}^2\rangle \langle W_{\rm i}\rangle -
   \langle W_{\rm s}\rangle \langle W_{\rm i}^2\rangle < 0, \hspace{3mm} \label{36}\\
 B^{40}_{31} &\equiv& \langle W_{\rm s}^4\rangle + \langle W_{\rm i}^4\rangle -
   \langle W_{\rm s}^3\rangle \langle W_{\rm i}\rangle -
   \langle W_{\rm s}\rangle \langle W_{\rm i}^3\rangle < 0, \label{37}\\
 B^{31}_{22} &\equiv& \langle W_{\rm s}^3\rangle \langle W_{\rm i}\rangle + \langle W_{\rm s}\rangle \langle W_{\rm i}^3\rangle
   - 2\langle W_{\rm s}^2\rangle \langle W_{\rm i}^2\rangle < 0, \label{38}\\
 B^{50}_{41} &\equiv& \langle W_{\rm s}^5\rangle + \langle W_{\rm i}^5\rangle -
   \langle W_{\rm s}^4\rangle \langle W_{\rm i}\rangle -
   \langle W_{\rm s}\rangle \langle W_{\rm i}^4\rangle < 0, \label{39}\\
 B^{41}_{32} &\equiv& \langle W_{\rm s}^4\rangle \langle W_{\rm i}\rangle + \langle W_{\rm s}\rangle \langle W_{\rm i}^4\rangle
   - \langle W_{\rm s}^3\rangle \langle W_{\rm i}^2\rangle \nonumber \\
 & & \mbox{} - \langle W_{\rm s}^2\rangle \langle W_{\rm i}^3\rangle < 0. \label{40}
\end{eqnarray}

The above LNCCa can be completed with simpler criteria that have
their origin in the majorization theory applied to polynomials
written in two independent variables $ W_a $ and $ W'_a $ that
uses averaging with the quasi-distribution function $
P_a(W_a)P_a(W'_a) $, $ a = {\rm s,i} $. These \emph{Local
NonClassicality Criteria} are obtained in the form:
\cite{Lee1990a,Verma2010,Arkhipov2016c,PerinaJr2017}:
\begin{eqnarray}  
 ^aL^{20}_{11} &\equiv& \langle W_a^2\rangle - \langle W_a\rangle^2 <0, \label{41}\\
 ^aL^{30}_{21} &\equiv& \langle W_a^3\rangle - \langle W_a^2\rangle \langle W_a\rangle < 0, \label{42}\\
 ^aL^{40}_{31} &\equiv& \langle W_a^4\rangle - \langle W_a^3\rangle \langle W_a\rangle < 0, \label{43}\\
 ^aL^{31}_{22} &\equiv& \langle W_a^3\rangle \langle W_a\rangle - \langle W_a^2\rangle^2 < 0, \label{44}\\
 ^aL^{50}_{41} &\equiv& \langle W_a^5\rangle - \langle W_a^4\rangle \langle W_a\rangle < 0, \label{45}\\
 ^aL^{41}_{32} &\equiv& \langle W_a^4\rangle \langle W_a\rangle - \langle W_a^3\rangle \langle W_a^2\rangle < 0. \label{46}
\end{eqnarray}
We note that the LNCCa given in Eqs.~(\ref{35})---(\ref{46}) occur
in more complex expressions derived below, that combine the local
nonclassicalities with the entanglement. We also note that the
simplest LNCC given in Eq.~(\ref{41}) was experimentally observed
already in 1977 using the light from fluorescence of a single
molecule \cite{Kimble1977}.

To reveal more complex GNCCa, we first analyze the formulas of the
majorization theory with three independent variables $ W_{\rm s}
$, $ W_{\rm i} $ and $ W'_{a} $ considering two kinds of averaging
with the quasi-distribution functions $ P_{\rm si}(W_{\rm
s},W_{\rm i}) P_a(W'_a) $, $ a={\rm s,i} $. To demonstrate the
structure of the obtained inequalities without treating more
complex formulas, we investigate the inequalities including
intensity moments up to the fourth order. Detailed analysis of the
majorization formulas denoted in the standard notation as $
\{200\} \succ \{110\} $, $ \{300\} \succ \{210\} $, $ \{400\}
\succ \{310\} $, and $ \{310\} \succ \{220\} $ reveals that all
these inequalities are obtained as suitable positive linear
combinations of some of the inequalities written in
Eqs.~(\ref{20})---(\ref{29}) and (\ref{35})---(\ref{46}) and so
they are redundant for the indication of nonclassicality. They can
be found in Appendix A [Eqs.~(\ref{A17})---(\ref{A20})]. The
remaining majorization inequalities $ \{210\} \succ \{111\} $ and
$ \{220\} \succ \{211\} $ considered with both types of averaging
then provide the following four \emph{Global NonClassicality
Criteria} ($ a={\rm s,i} $):
\begin{eqnarray}  
 ^{a}D^{210}_{111} &\equiv& 2 \langle W_{a}^2\rangle\langle W_{a}\rangle +
  \langle W_{\rm s}^2W_{\rm i}\rangle + \langle W_{\rm s}W_{\rm i}^2\rangle
  + \langle W_{\rm s}^2\rangle\langle W_{\rm i}\rangle \nonumber \\
  & & \hspace{-3mm} \mbox{} + \langle W_{\rm s}\rangle\langle W_{\rm i}^2\rangle - 6\langle W_{a}\rangle
   \langle W_{\rm s}W_{\rm i}\rangle <0, \label{47}\\
 ^{a}D^{220}_{211} &\equiv& \langle W_{\rm a}^2\rangle^2 +
  \langle W_{\rm s}^2W_{\rm i}^2\rangle + \langle W_{\rm s}^2\rangle\langle W_{\rm i}^2\rangle
  - \langle W_{a}\rangle \nonumber \\
  & & \hspace{-3mm}  \mbox{} \times [\langle W_{\rm s}^2W_{\rm i}\rangle +
   \langle W_{\rm s}W_{\rm i}^2\rangle] - \langle W_{a}^2\rangle
   \langle W_{\rm s}W_{\rm i}\rangle <0. \label{48}
\end{eqnarray}

In the next step, we analyze the majorization inequalities with
four independent variables $ W_{\rm s} $, $ W_{\rm i} $, $ W'_{\rm
s} $, and $ W'_{\rm i} $ and we use the quasi-distribution
function $ P_{\rm si}(W_{\rm s},W_{\rm i}) P_{\rm si}(W'_{\rm
s},W'_{\rm i}) $ for averaging. The inequalities $ \{2000\} \succ
\{1100\} $, $ \{3000\} \succ \{2100\} $, $ \{4000\} \succ \{3100\}
$, and $ \{3100\} \succ \{2200\} $ can be expressed as positive
linear combinations of those given in Eqs.(\ref{20})---(\ref{29})
and (\ref{35})---(\ref{46}) and as such they are not interesting
for revealing nonclassicality. Similarly, the doubled inequality $
\{2100\} \succ \{1110\} $ [$ \{2200\} \succ \{2110\} $] is
obtained as the sum $ ^{s}D^{210}_{111} + ^{i}D^{210}_{111} $ [$
^{s}D^{220}_{211} + ^{i}D^{220}_{211} $] of the GNCCa written in
Eq.~(\ref{47}) [(\ref{48})]. More details are given in Appendix~A
[see Eqs.~(\ref{A21})---(\ref{A26})]. Only the inequality $
\{2110\} \succ \{1111\} $ is recast into the following
\emph{Global NonClassicality Criterion:}
\begin{eqnarray}  
 D^{2110}_{1111} &\equiv& [\langle W_{\rm s}^2W_{\rm i}\rangle +
  \langle W_{\rm s}W_{\rm i}^2\rangle] [\langle W_{\rm s}\rangle
  + \langle W_{\rm i}\rangle ] + \langle W_{\rm s}W_{\rm i}\rangle \nonumber \\
  & & \mbox{} \times [ \langle W_{\rm s}^2\rangle + \langle W_{\rm i}^2\rangle]
    - 6\langle W_{\rm s}W_{\rm i}\rangle^2 <0.
\label{49}
\end{eqnarray}

The remaining inequalities up to the fourth order are provided by
the majorization inequalities $ \{2100\} \succ \{1110\} $, $
\{2200\} \succ \{2110\} $ and $ \{2110\} \succ \{1111\} $ if we
perform in turn averaging with the following three
quasi-distribution functions $ P_{\rm si}(W_{\rm s},W_{\rm i})
P_a(W'_{a})P_a(W''_{a}) $, $ a={\rm s,i} $, and $ P_{\rm
si}(W_{\rm s},W_{\rm i}) P_{\rm s}(W'_{\rm s}) P_{\rm i}(W'_{\rm
i}) $. The occurrence of three intensity moments in a product
represents their common feature. Step by step, the corresponding
\emph{Global NonClassicality Criteria} are derived in the form ($
a={\rm s,i} $):
\begin{eqnarray}  
 ^{a}T^{2100}_{1110} &\equiv& 6 \langle W_{a}^2\rangle\langle W_{a}\rangle +
  \langle W_{\rm s}^2W_{\rm i}\rangle + \langle W_{\rm s}W_{\rm i}^2\rangle
  + 2 \langle W_{\rm s}^2\rangle\langle W_{\rm i}\rangle \nonumber \\
  & & \mbox{} + 2 \langle W_{\rm s}\rangle\langle W_{\rm i}^2\rangle - 6\langle W_{a}\rangle
   \langle W_{\rm s}W_{\rm i}\rangle \nonumber \\
  & & \mbox{} - 3\langle W_{a}\rangle^2 [\langle W_{\rm s}\rangle +
   \langle W_{\rm i}\rangle] <0, \label{50} \\
 T^{2100}_{1110} &\equiv& 2 \langle W_{\rm s}^2\rangle\langle W_{\rm s}\rangle +
   2 \langle W_{\rm i}^2\rangle\langle W_{\rm i}\rangle +
  \langle W_{\rm s}^2W_{\rm i}\rangle + \langle W_{\rm s}W_{\rm i}^2\rangle \nonumber \\
  & & \mbox{} + 3\langle W_{\rm s}^2\rangle\langle W_{\rm i}\rangle
   + 3\langle W_{\rm s}\rangle\langle W_{\rm i}^2\rangle -
   3[\langle W_{\rm s}\rangle + \langle W_{\rm i}\rangle] \nonumber \\
  & & \mbox{} \times \langle W_{\rm s}W_{\rm i}\rangle - 3\langle W_{\rm s}\rangle^2
   \langle W_{\rm i}\rangle - 3\langle W_{\rm s}\rangle
   \langle W_{\rm i}\rangle^2 <0, \nonumber \\
 & &  \label{51} \\
 {}^{a}T^{2200}_{2110} &\equiv& 6 \langle W_{a}^2\rangle^2 +
   2\langle W_{\rm s}^2W_{\rm i}^2\rangle + 4\langle W_{\rm s}^2\rangle\langle W_{\rm i}^2\rangle
   \nonumber \\
  & & \mbox{} - 2\langle W_{a}\rangle^2\langle W_{a}^2\rangle
   - \langle W_{a}\rangle^2 [\langle W_{\rm s}^2\rangle + \langle W_{\rm i}^2\rangle]
   \nonumber \\
  & & \mbox{}
   - 2\langle W_{a}\rangle [\langle W_{\rm s}^2W_{\rm i}\rangle +
   \langle W_{\rm s}W_{\rm i}^2\rangle] \nonumber \\
  & & \mbox{} - 2\langle W_{a}^2\rangle[\langle W_{\rm s}W_{\rm i}\rangle
   +\langle W_{\rm s}\rangle\langle W_{\rm i}\rangle ] <0, \label{52} \\
 T^{2200}_{2110} &\equiv& 2\langle W_{\rm s}^2\rangle^2 + 2\langle W_{\rm i}^2\rangle^2
   + 2\langle W_{\rm s}^2W_{\rm i}^2\rangle + 6\langle W_{\rm s}^2\rangle\langle W_{\rm i}^2\rangle
   \nonumber \\
  & & \mbox{} - [\langle W_{\rm s}\rangle + \langle W_{\rm i}\rangle]
   [\langle W_{\rm s}^2W_{\rm i}\rangle + \langle W_{\rm s}W_{\rm i}^2\rangle] \nonumber \\
  & & \mbox{} - [\langle W_{\rm s}^2\rangle + \langle W_{\rm i}^2\rangle]
   [\langle W_{\rm s}W_{\rm i}\rangle + 2\langle W_{\rm s}\rangle\langle W_{\rm i}\rangle ] \nonumber \\
  & & \mbox{} - \langle W_{\rm s}^2\rangle\langle W_{\rm i}\rangle^2
   - \langle W_{\rm s}\rangle^2\langle W_{\rm i}^2\rangle <0,
   \label{53} \\
 ^{a}T^{2110}_{1111} &\equiv& 2\langle W_{a}^2\rangle\langle W_{a}\rangle^2
   + 2\langle W_{a}\rangle [\langle W_{\rm s}^2W_{\rm i}\rangle +
   \langle W_{\rm s}W_{\rm i}^2\rangle] \nonumber \\
  & & \mbox{} + \langle W_{a}\rangle^2 [ \langle W_{\rm s}^2\rangle + \langle W_{\rm i}^2\rangle]
   \nonumber \\
  & & \mbox{} + 2\langle W_{a}^2\rangle[\langle W_{\rm s}W_{\rm i}\rangle
   +\langle W_{\rm s}\rangle\langle W_{\rm i}\rangle ] \nonumber\\
  & & \mbox{} - 12\langle W_{a}\rangle^2\langle W_{\rm s}W_{\rm i}\rangle
   <0, \label{54} \\
 T^{2110}_{1111} &\equiv& [\langle W_{\rm s}\rangle + \langle W_{\rm i}\rangle]
   [\langle W_{\rm s}^2W_{\rm i}\rangle + \langle W_{\rm s}W_{\rm i}^2\rangle] \nonumber \\
  & & \mbox{} + [\langle W_{\rm s}^2\rangle + \langle W_{\rm i}^2\rangle]
   [\langle W_{\rm s}W_{\rm i}\rangle + 2\langle W_{\rm s}\rangle\langle W_{\rm i}\rangle ] \nonumber \\
  & & \mbox{} + \langle W_{\rm s}^2\rangle\langle W_{\rm i}\rangle^2
   + \langle W_{\rm s}\rangle^2\langle W_{\rm i}^2\rangle \nonumber\\
  & & \mbox{} - 12 \langle W_{\rm s}\rangle\langle W_{\rm i}\rangle \langle W_{\rm s}W_{\rm i}\rangle
   <0. \label{55}
\end{eqnarray}
We note that the approach leading to Eqs.~(\ref{50})---(\ref{55})
provides also additional redundant GNCCa that are summarized in
Appendix~A [see Eqs.~(\ref{A27})---(\ref{A34})].

Additional nonclassicality inequalities containing products of
three intensity moments are reached from the majorization
inequalities written for polynomials with three variables and
assuming averaging with the factorized quasi-distributions $
P_{\rm s}(W_{\rm s})P_{\rm i}(W_{\rm i})P_{a}(W'_{a}) $, $ a={\rm
s,i} $. The majorization inequalities $ \{210\} \succ \{111\} $
and $ \{220\} \succ \{211\} $ leave us with the following
\emph{Local NonClassicality Criteria} in this case ($ a={\rm s,i}
$):
\begin{eqnarray}  
 ^{a}B^{210}_{111} &\equiv& \langle W_{a}^2\rangle\langle W_{a}\rangle
   + \langle W_{\rm s}^2\rangle\langle W_{\rm i}\rangle + \langle W_{\rm i}^2\rangle\langle W_{\rm s}\rangle
   \nonumber \\
  & & \mbox{} - 3\langle W_{a}\rangle\langle W_{\rm s}\rangle\langle W_{\rm i}\rangle < 0 , \label{56}\\
 ^{a}B^{220}_{211} &\equiv& \langle W_{a}^2\rangle^2 + 2\langle W_{\rm s}^2\rangle\langle W_{\rm i}^2\rangle
   + \langle W_{a}\rangle^2\langle W_{a}^2\rangle - \langle W_{a}\rangle^2 \nonumber \\
  & & \mbox{} \times [\langle W_{\rm s}^2\rangle +\langle W_{\rm i}^2\rangle]
   - 2\langle W_{a}^2\rangle \langle W_{\rm s}\rangle\langle W_{\rm i}\rangle < 0 . \label{57}
\end{eqnarray}

Analyzing the inequalities originating in the majorization theory
with intensity moments up to the fourth order, we finally arrive
at those written among the terms with four intensity moments in
the product. They are naturally derived from the majorization
inequalities written for polynomials with four variables
considering in turn the quasi-distributions $ P_{\rm s}(W_{\rm
s})P_{\rm i}(W_{\rm i})P_{a}(W'_{a})P_{a}(W''_{a}) $, $ a={\rm
s,i} $, and $ P_{\rm s}(W_{\rm s})P_{\rm i}(W_{\rm i})P_{\rm
s}(W'_{\rm s})P_{\rm i}(W'_{\rm i}) $. In detail, the majorization
inequality $ \{2110\} \succ \{1111\} $ is recast considering the
above averaging into the following \emph{Local NonClassicality
Criteria} ($ a={\rm s,i} $):
\begin{eqnarray}  
 ^{a}B^{2110}_{1111} &\equiv& \langle W_{a}\rangle^2 [\langle W_{\rm s}^2\rangle
   + \langle W_{\rm i}^2\rangle] + 2\langle W_{a}^2\rangle\langle W_{\rm s}\rangle\langle W_{\rm i}\rangle
   \nonumber \\
  & & \mbox{} -4 \langle W_{a}\rangle^2 \langle W_{\rm s}\rangle\langle W_{\rm i}\rangle < 0 , \label{58}\\
 B^{2110}_{1111} &\equiv& \langle W_{\rm s}^2\rangle\langle W_{\rm i}\rangle^2 +
  \langle W_{\rm s}\rangle^2\langle W_{\rm i}^2\rangle + 2[\langle W_{\rm s}^2\rangle
   + \langle W_{\rm i}^2\rangle] \nonumber \\
  & & \mbox{} \times \langle W_{\rm s}\rangle\langle W_{\rm
   i}\rangle - 6\langle W_{\rm s}\rangle^2\langle W_{\rm i}\rangle^2 < 0 . \label{59}
\end{eqnarray}
We note that also additional LNCCa arise from the majorization
theory written for polynomials with three and four variables.
However, they can be expressed as positive linear combinations of
the above written LNCCa and so they are redundant. They are
explicitly given in Eqs.~(\ref{A1})---(\ref{A16}) in Appendix~A.

\subsection{Nonclassicality criteria based on non-negative polynomials}

Similarly as in the previous section where we have used the mean
values of non-negative polynomials in Eq.~(\ref{19}), here we
derive \emph{Local} and \emph{Global NonClassicality Criteria} by
negating the following classical inequalities:
\begin{eqnarray}  
 \langle W_{\rm s}^k W_{\rm i}^l (W_{\rm s}-\langle W_{\rm s}\rangle)^{2m}
  (W_{\rm i}-\langle W_{\rm i}\rangle)^{2n}\rangle &<&0, \nonumber \\
 & & \hspace{-40mm} k,l=0,1,\ldots, \;\; m,n=0,1\ldots.
\label{60}
\end{eqnarray}

Concentrating on the signal field ($ m=1 $ and $n=0 $) and
restricting our attention to the LNCCa containing intensity
moments up to the fifth order we recognize in Eqs.~(\ref{60}) the
following LNCCa:
\begin{eqnarray}  
 E_{0l10} &\equiv& \langle W_{\rm s}^2W_{\rm i}^l\rangle + \langle W_{\rm s}\rangle^2 \langle W_{\rm i}^l\rangle
   - 2\langle W_{\rm s}\rangle \langle W_{\rm s}W_{\rm i}^l\rangle <0, \nonumber \\
  & &  \hspace{10mm} l=1,2,3, \label{61} \\
 E_{1l10} &\equiv& \langle W_{\rm s}^3W_{\rm i}^l\rangle + \langle W_{\rm s}\rangle^2 \langle W_{\rm s}W_{\rm i}^l\rangle
   - 2\langle W_{\rm s}\rangle \langle W_{\rm s}^2W_{\rm i}^l\rangle <0, \nonumber \\
  & &  \hspace{10mm} l= 1,2, \label{62} \\
 E_{2110} &\equiv& \langle W_{\rm s}^4W_{\rm i}\rangle + \langle W_{\rm s}\rangle^2 \langle W_{\rm s}^2W_{\rm i}\rangle
   - 2\langle W_{\rm s}\rangle \langle W_{\rm s}^3W_{\rm i}\rangle <0.  \nonumber \\
  & &\label{63}
\end{eqnarray}
One additional LNCC ($ E_{0120} $) as well as one additional GNCC
($ E_{1011} $) are expressed as linear combinations of the LNCCa
in Eqs.~(\ref{61})---(\ref{63}) with varying signs:
\begin{eqnarray}  
 E_{0120} &\equiv&  E_{2110} + \langle W_{\rm s}\rangle^2 E_{0110}
  - 2\langle W_{\rm s}\rangle E_{1110}<0, \label{64} \\
 E_{1011} &\equiv&  E_{1210} + \langle W_{\rm i}\rangle^2 E_{1010}
  - 2\langle W_{\rm i}\rangle E_{1110}<0. \label{65}
\end{eqnarray}
The LNCCa and GNCC given in Eqs.(\ref{61})---(\ref{65}) with
exchanged subscripts s and i provide additional LNCCa and GNCC
that can be derived from the symmetry. Moreover, there exists
another GNCC belonging to the fourth order and being symmetric
with respect to subscripts s and i:
\begin{eqnarray}  
 E_{0011} \equiv E_{0210} + \langle W_{\rm i}\rangle^2\;{}^{\rm s}L^{20}_{11}
  - 2\langle W_{\rm i}\rangle E_{0110}<0. \label{66}
\end{eqnarray}

We note that Eq.~(\ref{60}) considered for $ l=n=0 $ gives also
nontrivial LNCCa that can be added to those written in
Eqs.~(\ref{41})---(\ref{46}). They are expressed as:
\begin{eqnarray}   
 E_{1010} &\equiv& {}^{\rm s}L^{30}_{21} - \langle W_{\rm s}\rangle \; {}^{\rm
  s}L^{20}_{11} < 0, \label{67} \\
 E_{2010} &\equiv& {}^{\rm s}L^{40}_{31} - \langle W_{\rm s}\rangle \; {}^{\rm
  s}L^{30}_{21} < 0, \label{68} \\
 E_{3010} &\equiv& {}^{\rm s}L^{50}_{41} - \langle W_{\rm s}\rangle \; {}^{\rm
  s}L^{40}_{31} < 0, \label{69} \\
 E_{0020} &\equiv& {}^{\rm s}L^{40}_{31} - 3\langle W_{\rm s}\rangle \; {}^{\rm
  s}L^{30}_{21} + 3\langle W_{\rm s}\rangle^2 \; {}^{\rm s}L^{20}_{11} < 0, \label{70} \\
 E_{1020} &\equiv& {}^{\rm s}L^{50}_{41} - 3\langle W_{\rm s}\rangle \;{}^{\rm
  s}L^{40}_{31} + 3\langle W_{\rm s}\rangle^2 \;{}^{\rm s}L^{30}_{21} \nonumber \\
 & & \mbox{} - \langle W_{\rm s}\rangle^3 \;{}^{\rm s}L^{20}_{11} < 0. \label{71}
\end{eqnarray}

\subsection{Global nonclassicality criteria based on the matrix approach}

In this case, the GNCCa are based on considering classically
positive semi-definite matrices of dimension $ n\times n $ for $
n=2,3,\ldots $ that describe mean values of quadratic forms
defined above the basis that includes different powers of the
signal and idler intensities. This approach has been elaborated in
general both for the amplitude and intensity moments in Refs.
\cite{Agarwal1992,Shchukin2005,Miranowicz2006,Vogel2008},
summarized in Ref. \cite{Miranowicz2010} and applied in Ref.
\cite{Sperling2015}. The Bochner theorem has been used to arrive
at the even more general forms of these inequalities
\cite{Richter2002,Ryl2015}. For $ n=2 $ the \emph{Global
NonClassicality Criteria} are defined along the relation ($
i,j,k,l \ge 0 $):
\begin{eqnarray}  
 & M_{ijkl} \equiv \langle W_{\rm s}^{2i}W_{\rm i}^{2j}\rangle
  \langle W_{\rm s}^{2k}W_{\rm i}^{2l}\rangle
  - \langle W_{\rm s}^{i+k}W_{\rm i}^{j+l}\rangle^2 < 0. \nonumber
  \\
 & &
\label{72}
\end{eqnarray}

Restricting our considerations to the GNCCa up to the fifth order
in intensity moments, we only reveal the following two
inequalities:
\begin{eqnarray}  
 M_{1100} &\equiv& \langle W_{\rm s}^{2}W_{\rm i}^{2}\rangle
  - \langle W_{\rm s}W_{\rm i}\rangle^2 < 0, \label{73} \\
 M_{1001} &\equiv& \langle W_{\rm s}^{2}\rangle\langle W_{\rm i}^{2}\rangle
  - \langle W_{\rm s}W_{\rm i}\rangle^2 < 0. \label{74}
\end{eqnarray}

For comparison, we write down two GNCCa originating in the
majorization inequalities $ \{2200\} \succ \{1111\} $ and $
\{4000\} \succ \{1111\} $ considered with averaging over the
quasi-distribution function $ P_{\rm si}(W_{\rm s},W_{\rm i})
P_{\rm si}(W'_{\rm s},W'_{\rm i}) $:
\begin{eqnarray}  
 D^{2200}_{1111} &\equiv& [\langle W_{\rm s}^{2}\rangle
  + \langle W_{\rm i}^{2}\rangle]^2 + 2 \langle W_{\rm s}^{2}W_{\rm i}^{2}\rangle
  - 6\langle W_{\rm s}W_{\rm i}\rangle^2 < 0, \nonumber \\
 & &  \label{75} \\
 D^{4000}_{1111} &\equiv& \langle W_{\rm s}^4\rangle
  + \langle W_{\rm i}^4\rangle - 2\langle W_{\rm s}W_{\rm i}\rangle^2 < 0. \label{76}
\end{eqnarray}
We note that the GNCCa $ D^{2200}_{1111} $ and $ D^{4000}_{1111} $
stem from the GNCCa written in Eqs.~(\ref{47})---(\ref{49}) and
the LNCCa summarized in Eqs.~(\ref{35})---(\ref{46}).

Also a $ 3\times 3 $ matrix built above the base vector $ (1,
W_{\rm s}, W_{\rm i}) $ results in one \emph{Global
NonClassicality Criterion} of the fourth order:
\begin{eqnarray}  
 M_{001001} &\equiv& \langle W_{\rm s}^{2}\rangle\langle W_{\rm i}^{2}\rangle
   + 2\langle W_{\rm s}W_{\rm i}\rangle\langle W_{\rm s}\rangle\langle W_{\rm i}\rangle
   - \langle W_{\rm s}W_{\rm i}\rangle^2 \nonumber \\
  & & \mbox{} - \langle W_{\rm s}^2\rangle\langle W_{\rm i}\rangle^2
   - \langle W_{\rm s}\rangle^2\langle W_{\rm i}^2\rangle < 0.
   \label{77}
\end{eqnarray}

\subsection{Global nonclassicality criteria derived from the Cauchy-Schwarz inequality}

To reveal additional \emph{Global NonClassicality Criteria}, we
negate the Cauchy-Schwarz inequality:
\begin{eqnarray}  
 &\left[ \int dW_{\rm s}dW_{\rm i} P_{\rm si}(W_{\rm s},W_{\rm i})
   f(W_{\rm s},W_{\rm i}) g(W_{\rm s},W_{\rm i})\right]^2
   \nonumber \\
 & \hspace{5mm}  > \int dW_{\rm s}dW_{\rm i} P_{\rm si}(W_{\rm s},W_{\rm i})
   f^2(W_{\rm s},W_{\rm i}) \nonumber \\
 & \hspace{5mm} \mbox{} \times
 \int dW_{\rm s}dW_{\rm i} P_{\rm si}(W_{\rm s},W_{\rm i})
   g^2(W_{\rm s},W_{\rm i}).
\label{78}
\end{eqnarray}
In Eq.~(\ref{78}), $ f $ and $ g $ denote arbitrary real functions
and $ P_{\rm si} $ stands for the joint quasi-distribution of
integrated intensities. Restricting ourselves up to the fifth
power of intensities, we may in turn consider $ f = 1 $ together
with $ g = W_{\rm s}W_{\rm i} $, $ f= \sqrt{W_{\rm s}} $ together
with $ g = \sqrt{W_{\rm s}} W_{\rm i} $, $ f= W_{\rm s} $ together
with $ g = W_{\rm i} $, and $ f= W_{\rm s}\sqrt{W_{\rm i}} $
together with $ g = \sqrt{W_{\rm i}}$ to arrive at the following
GNCCa:
\begin{eqnarray}  
 C^{00}_{22} &\equiv& \langle W_{\rm s}^2W_{\rm i}^2\rangle -
 \langle W_{\rm s}W_{\rm i}\rangle^2 < 0, \label{79} \\
 C^{10}_{12} &\equiv& \langle W_{\rm s}W_{\rm i}^2\rangle  \langle W_{\rm s}\rangle -
 \langle W_{\rm s}W_{\rm i}\rangle^2 < 0, \label{80} \\
 C^{20}_{02} &\equiv& \langle W_{\rm s}^2\rangle \langle W_{\rm i}^2\rangle -
 \langle W_{\rm s}W_{\rm i}\rangle^2 < 0, \label{81} \\
 C^{21}_{01} &\equiv& \langle W_{\rm s}^2W_{\rm i}\rangle \langle W_{\rm i}\rangle -
 \langle W_{\rm s}W_{\rm i}\rangle^2 < 0. \label{82}
\end{eqnarray}
The criterion $ C^{00}_{22} $ in Eq.~(\ref{79}) [$ C^{20}_{02} $
in Eq.~(\ref{81})] coincides with the criterion $ M_{1100} $ in
Eq.~(\ref{73}) [$ M_{1001} $ in Eq.~(\ref{74})] derived from the
matrix approach.

All inequalities among the intensity moments discussed both in the
previous and this section can mutually be compared quantitatively
when we transform these inequalities into the corresponding
nonclassicality depths. In this approach, we replace the usual
(normally-ordered) intensity moments $ \langle W^k\rangle $ by
moments $ \langle W^k\rangle_s $ related to a general $ s $
ordering of the field operators according to the formula
\cite{Perina1991}
\begin{equation}   
 \langle W^k\rangle_s = \left(\frac{2}{1-s}\right)^k \left\langle
  {\rm L}_k \left(\frac{2W}{s-1}\right) \right\rangle
\label{83}
\end{equation}
in which $ {\rm L}_k $ denotes the $ k $-th Laguerre polynomial
\cite{Gradshtein2000}. Then, we formally consider all the above
inequalities originally derived for the normally-ordered intensity
moments with $ s $-ordered intensity moments and varying value of
the parameter $ s $. If a given inequality indicates
nonclassicality for the normally-ordered moments, decreasing
values of the ordering parameter $ s $ gradually suppress this
nonclassicality due to the increasing additional 'detection' noise
\cite{Lee1991}. The nonclassicality is lost for certain threshold
value $ s_{\rm th} $. This value defines a nonclassicality depth
(NCD) $ \tau $ \cite{Lee1991} as follows:
\begin{equation}    
 \tau = \frac{1-s_{\rm th} }{2} .
\label{84}
\end{equation}
The greater the value of NCD $ \tau $ is the stronger the
nonclassicality is.

\section{Nonclassicality criteria based on the elements of photocount and photon-number distributions and
 their moments}

All nonclassicality criteria based on intensity moments and widely
discussed in the previous two sections can be easily transformed
into the corresponding criteria that use the elements of
photon-number [photocount] distribution $ p_{\rm si}(n_{\rm
s},n_{\rm i}) $ [$ f_{\rm si}(c_{\rm s},c_{\rm i}) $]
\cite{Klyshko1996,Waks2004,Waks2006,PerinaJr2017}. To understand
this, we first write down the two-dimensional Mandel
photodetection formula \cite{Perina1991,Mandel1995}:
\begin{eqnarray}   
 p_{\rm si}(n_{\rm s},n_{\rm i}) \hspace{-0.5mm}&=& \hspace{-0.5mm} \frac{1}{n_{\rm s}!\, n_{\rm i}!}
  \int_{0}^{\infty} dW_{\rm s}\int_{0}^{\infty} dW_{\rm i}\,
  W_{\rm s}^{n_{\rm s}} W_{\rm i}^{n_{\rm i}} \nonumber \\
 & & \mbox{} \times \exp[-(W_{\rm s}+W_{\rm i})] P_{\rm si}(W_{\rm s},W_{\rm i}),
\label{85}
\end{eqnarray}
where $ P_{\rm si}(W_{\rm s},W_{\rm i}) $ is the above used joint
quasi-distribution of integrated intensities. Introducing the
modified elements $ \tilde p_{\rm si} $ of the photon-number
distribution,
\begin{equation} 
 \tilde p_{\rm si}(n_{\rm s},n_{\rm i}) \equiv \frac{ n_{\rm s}!\,
  n_{\rm i}!\, p_{\rm si}(n_{\rm s},n_{\rm i}) }{p_{\rm si}(0,0)} ,
\label{86}
\end{equation}
and the properly normalized quasi-distribution $ \tilde P_{\rm si}
$,
\begin{eqnarray} 
 \tilde P_{\rm si}(W_{\rm s},W_{\rm i}) \hspace{-0.5mm}&\equiv&\hspace{-0.5mm}
   \exp[-(W_{\rm s}+W_{\rm i})] P_{\rm si}(W_{\rm s},W_{\rm i}) \nonumber \\
 & & \hspace{-22mm} \times \left[ \int_{0}^{\infty} dW_{\rm s}\int_{0}^{\infty} dW_{\rm i}
  \exp[-(W_{\rm s}+W_{\rm i})] P_{\rm si}(W_{\rm s},W_{\rm i})
  \right]^{-1}, \nonumber \\
 & &
\label{87}
\end{eqnarray}
the Mandel photodetection formula in Eq.~(\ref{85}) is recast into
the form defining the modified elements $ \tilde p_{\rm si} $ as
the moments of the quasi-distribution $ \tilde P_{\rm si} $:
\begin{equation}   
 \tilde p_{\rm si}(n_{\rm s},n_{\rm i}) = \int_{0}^{\infty} dW_{\rm s}\int_{0}^{\infty} dW_{\rm i}
  \, W_{\rm s}^{n_{\rm s}} W_{\rm i}^{n_{\rm i}} \tilde P_{\rm si}(W_{\rm s},W_{\rm
  i}).
\label{88}
\end{equation}
The formal substitution in the above derived nonclassicality
criteria for intensity moments suggested by formula (\ref{88}) is
expressed as
\begin{equation}  
 \langle W_{\rm s}^{n_{\rm s}} W_{\rm i}^{n_{\rm i}} \rangle
 \longleftarrow \tilde p_{\rm si}(n_{\rm s},n_{\rm i}).
\label{89}
\end{equation}

As an example, we rewrite the inequalities in Eq.~(\ref{19}) for $
m=1 $ into the following \emph{Global NonClassicality Criteria}:
\begin{eqnarray}  
 F_{kl1}&\equiv& \tilde p_{\rm si}(k+2,l) + \tilde p_{\rm si}(k,l+2)
  - 2\tilde p_{\rm si}(k+1,l+1) < 0 , \nonumber \\
 & & \hspace{10mm} k,l = 0,1,\ldots.
\label{90}
\end{eqnarray}

Alternatively, the inequalities for intensity moments can be
directly transformed into the moments of photon numbers
(photocounts) exploiting the relation between the 'factorial'
photon-number moments (intensity moments) $ \langle W^k\rangle $
and usual photon-number moments $ \langle n^k\rangle $. Using the
Stirling numbers $ S(k,l) $ of the second kind \cite{Verma2010},
its two-dimensional variant is expressed in the form:
\begin{eqnarray}  
 \langle n_{\rm s}^{k_{\rm s}} n_{\rm i}^{k_{\rm i}}\rangle
   \hspace{-0.5mm}&=&\hspace{-0.5mm} \sum_{l_{\rm s}=1}^{k_{\rm s}}
  S^{-1}(k_{\rm s},l_{\rm s}) \sum_{l_{\rm i}=1}^{k_{\rm i}}
  S^{-1}(k_{\rm i},l_{\rm i}) W_{\rm s}^{l_{\rm s}} W_{\rm i}^{l_{\rm i}} ,
  \nonumber \\
 & & \hspace{10mm} k_{\rm s},k_{\rm i} =1,2,\ldots.
\label{91}
\end{eqnarray}
The Stirling numbers $ S(k,l) $ of the second kind for the first
five moments are conveniently expressed as a matrix $ S_{kl} $
that, together with its inverse matrix $ S_{kl}^{-1} $ giving the
Stirling numbers of the first kind, take the form:
\begin{eqnarray}  
 S_{kl} &=& \left[ \begin{array}{ccccc}
  1 & 0 & 0 & 0 & 0 \\
  1 & 1 & 0 & 0 & 0 \\
  1 & 3 & 1 & 0 & 0 \\
  1 & 7 & 6 & 1 & 0 \\
  1 & 15 & 25 & 10 & 1 \end{array} \right],
 \nonumber \\
 S_{kl}^{-1} &=& \left[ \begin{array}{ccccc}
  1 & 0 & 0 & 0 & 0 \\
  -1 & 1 & 0 & 0 & 0 \\
  2 & -3 & 1 & 0 & 0 \\
  -6 & 11 & -6 & 1 & 0 \\
  24 & -50 & 35 & -10 & 1 \end{array} \right].
\label{92}
\end{eqnarray}
We note that the above formulas between the intensity and
photon-number moments assume an effective single mode field.
However, generalization to multi-mode fields may be considered, as
it has been done for multi-mode twin beams in
Refs.~\cite{Allevi2012,Allevi2012a}. Also, different LNCCa
expressed either in the intensity or photon-number moments have
been compared in \cite{PerinaJr2017}.

The linear relations between the photon-number moments and the
intensity moments formulated in Eq.~(\ref{91}) can be used to
rewrite the nonclassicality criteria from the previous two
sections in terms of the photon-number moments. This is
interesting as the joint photocount distributions are directly
experimentally accessible and the joint photon-number
distributions are reached once we correct the experimental data
for finite detection efficiencies \cite{PerinaJr2012}. The
rewritten nonclassicality criteria usually attain, however, more
complex forms compared to the original ones written for intensity
moments. For this reason, we derive here only the nonclassicality
criteria that involve cross-correlation moments containing
different powers of the signal and idler photon numbers. They are
obtained as suitable positive linear combinations of the GNCCa $ E
$ written in Eqs.~(\ref{20})---(\ref{29}):
\begin{eqnarray}  
 N_{11} &\equiv& E_{001} \nonumber \\
  &=& \sum_{a={\rm s,i}} [\langle n_a^2\rangle - \langle
   n_a\rangle] - 2\langle n_{\rm s}n_{\rm i}\rangle <0,
  \label{93}  \\
 N_{21} &\equiv& E_{101} + E_{011} + E_{001} \nonumber \\
  &=& \sum_{a={\rm s,i}} [\langle n_a^3\rangle -2\langle n_a^2\rangle
   + \langle n_a\rangle] - \langle n_{\rm s}^2n_{\rm i}\rangle
   - \langle n_{\rm s}n_{\rm i}^2\rangle <0, \nonumber \\
  & &  \label{94}  \\
 N_{31} &\equiv& E_{201} + E_{021} + E_{111} + 3(E_{101} + E_{011} + E_{001}) \nonumber \\
  &=& \sum_{a={\rm s,i}} [\langle n_a^4\rangle - 3\langle n_a^3\rangle + 5\langle n_a^2\rangle
   - 3\langle n_a\rangle] - 4\langle n_{\rm s}n_{\rm i}\rangle \nonumber \\
  & & \mbox{} - \langle n_{\rm s}^3n_{\rm i}\rangle
   - \langle n_{\rm s}n_{\rm i}^3\rangle <0, \label{95}  \\
 N_{22} &\equiv& E_{201} + E_{021} + 2E_{111} + 2(E_{101} + E_{011} + E_{001}) \nonumber \\
  &=& \sum_{a={\rm s,i}} [\langle n_a^4\rangle - 4\langle n_a^3\rangle + 7\langle n_a^2\rangle
   - 4\langle n_a\rangle] - 2\langle n_{\rm s}n_{\rm i}\rangle \nonumber \\
  & & \mbox{} - 2\langle n_{\rm s}^2n_{\rm i}^2\rangle <0, \label{96}  \\
 N_{41} &\equiv& E_{301} + E_{031} + E_{211} + E_{121} + 6(E_{201} + E_{021}
  \nonumber \\
  & & \mbox{}  + E_{111}) + 7(E_{101} + E_{011} + E_{001}) \nonumber \\
  &=& \sum_{a={\rm s,i}} [\langle n_a^5\rangle - 4\langle n_a^4\rangle + 6\langle n_a^3\rangle
   + 2\langle n_a^2\rangle - 5\langle n_a\rangle] \nonumber \\
  & & \mbox{}  - 12\langle n_{\rm s}n_{\rm i}\rangle - \langle n_{\rm s}^4n_{\rm i}\rangle
   - \langle n_{\rm s}n_{\rm i}^4\rangle <0, \label{97}  \\
 N_{32} &\equiv& E_{301} + E_{031} + 2E_{211} + 2E_{121} + 4(E_{201} + E_{021})
  \nonumber \\
  & & \mbox{}  + 7E_{111} + 4(E_{101} + E_{011}) + E_{001} \nonumber \\
  &=& \sum_{a={\rm s,i}} [\langle n_a^5\rangle - 6\langle n_a^4\rangle + 15\langle n_a^3\rangle
   - 17\langle n_a^2\rangle + 7\langle n_a\rangle] \nonumber \\
  & & \mbox{}  - \langle n_{\rm s}^3n_{\rm i}^2\rangle
   - \langle n_{\rm s}^2n_{\rm i}^3\rangle <0. \label{98}
\end{eqnarray}

\section{Experimental verification of the derived nonclassicality and entanglement
criteria}

\begin{figure}  
 \centerline{\resizebox{0.8\hsize}{!}{\includegraphics{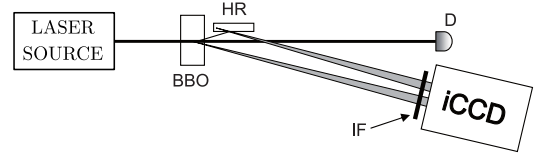}}}
 \caption{Scheme of the experimental setup:  A twin beam originating
 in a nonlinear crystal (BBO) pumped by an ultrashort pulse generates
 a weak twin beam. The signal field and the idler field (after reflection on mirror
 HR) are filtered by bandpass interference filter IF and then detected by an iCCD camera.
 The pump-beam intensity is actively stabilized by a feedback provided by detector D.}
\label{fig1}
\end{figure}

In order to experimentally judge the performance of the above
derived nonclassicality criteria, we have applied them to the
analysis of the entanglement between the signal and idler fields
constituting a weak twin beam generated in the process of
spontaneous parametric down-conversion
\cite{Mandel1995,PerinaJr2013a}. The marginal signal and idler
fields are generated with multi-mode thermal statistics which is a
consequence of the spontaneous emission. As such the twin beam is
locally classical and so the applied GNCCa are also the
entanglement criteria. The twin beam was generated in a 5-mm-long
type-I barium-borate crystal (BaB$ {}_2 $O$ {}_4 $, BBO) cut for a
slightly non-collinear geometry (for the experimental scheme, see
Fig.~\ref{fig1}). Parametric down-conversion was pumped by pulses
originating in the third harmonics (280~nm) of a femtosecond
cavity dumped Ti:sapphire laser that produced pulses with duration
150~fs and central wavelength 840~nm. The signal field as well as
the idler field were detected in different strips of the
photocathode of iCCD camera Andor DH334-18U-63. Before detection,
the nearly-frequency-degenerate signal and idler photons at the
wavelength of 560~nm were filtered by a 14-nm-wide bandpass
interference filter. Moreover, to stabilize the pump intensity,
and thus also the twin beam intensity, to minimize fluctuations in
the measured photocount distribution, the pump beam was actively
stabilized via a motorized half-wave plate followed by polarizer
and detector that monitored the actual intensity.

In the experiment, a joint signal-idler photocount histogram $
f_{\rm si}(c_{\rm s},c_{\rm i}) $ has been determined repeating
the measurement $ 1.2\times 10^6 $ times. This histogram obtained
with high precision due to the high number of repetitions has
allowed us to reconstruct the original joint signal-idler
photon-number distribution $ p_{\rm si}(n_{\rm s},n_{\rm i}) $
that characterizes the twin beam before being detected. We have
used two methods for making the reconstruction. First, we have
applied a method developed originally for detector calibration
\cite{PerinaJr2012a}. This method, in addition of giving the
detection efficiencies $ \eta_{\rm s} $ and $ \eta_{\rm i} $ in
the signal and idler fields, respectively, also gives parameters
of the used twin beam, though in a specific form of a multi-mode
Gaussian field. Knowing the detection efficiencies as well as
other parameters of the used iCCD camera, we have reconstructed
the measured twin beam by the general approach of expectation
maximization (maximum-likelihood approach) \cite{Dempster1977}.

\begin{figure}  
 \centerline{(a) \includegraphics[width=0.35\textwidth]{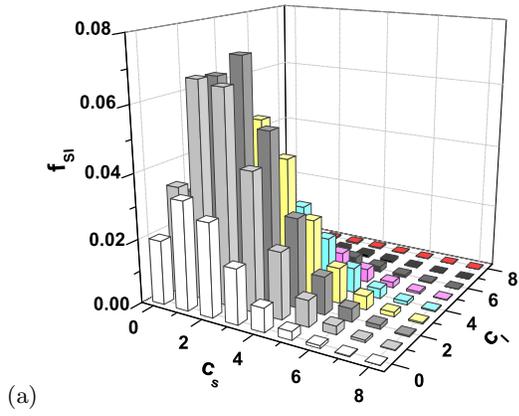}}
 \vspace{8mm}
 \centerline{(b) \includegraphics[width=0.35\textwidth]{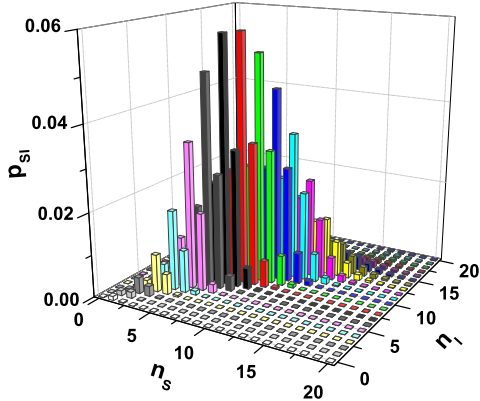}}
 \vspace{8mm}
 \centerline{(c) \includegraphics[width=0.35\textwidth]{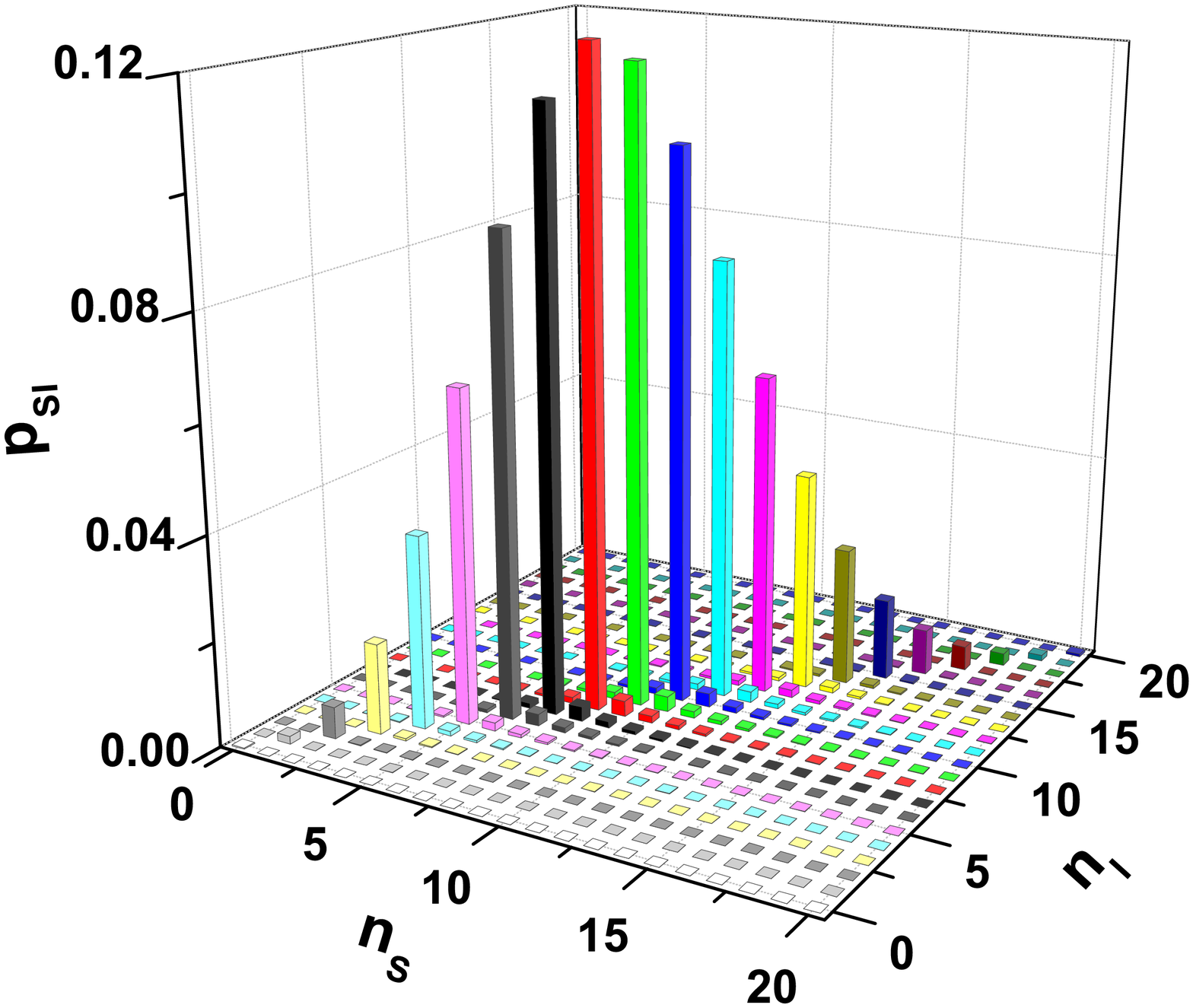}}
 \caption{(a) Experimental photocount histogram $ f_{\rm si}(c_{\rm s},c_{\rm i})$
  and reconstructed photon-number distributions $ p_{\rm si}(n_{\rm s},n_{\rm i})$
  obtained by (b) maximum-likelihood and (c) calibration
  methods.}
\label{fig2}
\end{figure}

In the calibration method, the twin beam has been revealed in the
analytical form of a multi-mode Gaussian field composed of
independent multi-mode paired, noise signal and noise idler
components characterized by mean photon(-pair) numbers $ B_a $ per
mode and numbers $ M_a $ of independent modes, $ a={\rm p,s,i} $
\cite{Perina2005,PerinaJr2013a}. The corresponding photon-number
distribution $ p_{\rm si}(n_{\rm s},n_{\rm i}) $ attains in this
case the form of a two-fold convolution among three Mandel-Rice
photon-number distributions \cite{Perina1991} belonging to the
constituting paired, noise signal and noise idler components
\cite{PerinaJr2012a,PerinaJr2013a,Perina2005}:
\begin{eqnarray}  
 p_{\rm si}(n_{\rm s},n_{\rm i}) &=& \sum_{n=0}^{{\rm min}[n_{\rm s},n_{\rm i}]}
  p(n_{\rm s}-n;M_{\rm s},B_{\rm s})
  p(n_{\rm i}-n;M_{\rm i},B_{\rm i}) \nonumber \\
 & & \mbox{} \times  p(n;M_{\rm p},B_{\rm p}).
\label{99}
\end{eqnarray}
The Mandel-Rice distribution $ p(n;M,B) $ is given as $ p(n;M,B) =
\Gamma(n+M) / [n!\, \Gamma(M)] B^n/(1+B)^{n+M} $ using the $
\Gamma $ function. Moreover, response of the iCCD camera has to be
described by an appropriate positive-operator-valued measure
(POVM). For an iCCD camera with $ N_a $ active pixels, detection
efficiency $ \eta_a $ and mean dark count number per pixel $ D_a
$, this POVM denoted as $ T_{\rm a}(c_{\rm a},n_{\rm a}) $ has
been derived in Ref.~\cite{PerinaJr2012}:
\begin{eqnarray}     
  T_a(c_a,n_a) &=& \left(\begin{array}{c} N_a \\ c_a \end{array}\right) (1-D_a)^{N_a}
   (1-\eta_a)^{n_a} (-1)^{c_a} \nonumber \\
  & &  \mbox{} \hspace{-18mm} \times  \sum_{l=0}^{c_a} \left(\begin{array}{c} c_a \\ l \end{array}\right)
    \frac{(-1)^l}{(1-D_a)^l}  \left( 1 + \frac{l}{N_a} \frac{\eta_a}{1-\eta_a}
   \right)^{n_a}. 
\label{100}
\end{eqnarray}
We note that the POVM  $ T_{\rm a}(c_{\rm a},n_{\rm a}) $ gives
the probability of having $ c_{\rm a} $ photocounts when detecting
a field with $ n_{\rm a} $ photons, $ a={\rm s,i} $. With these
premises, the method of the least squared declinations based on
the distribution $ p_{\rm si} $ in Eq.~(\ref{99}) and POVMs $
T_{\rm s} $ and $ T_{\rm i} $ for the signal and idler detection
arm, respectively, gives both the detection efficiencies $
\eta_{\rm s} $ and $ \eta_{\rm i} $ and parameters of the used
twin beam. The calibration method applied to the experimental
photocount histogram $ f_{\rm si}(c_{\rm s},c_{\rm i}) $ gave us
the following values of parameters: $ \eta_{\rm s}=0.230\pm 0.005
$, $ \eta_{\rm i}=0.220\pm 0.005 $, $M_{\rm p}=270$, $B_{\rm
p}=0.032$, $M_{\rm s}=0.01$, $B_{\rm s}=7.6$, $M_{\rm i} = 0.026$,
and $B_{\rm i}=5.3 $ (relative experimental errors: 7\%, for
details, see \cite{PerinaJr2013a}), in addition to those
determined independently: $N_{\rm s}=6528$, $N_{\rm i}=6784$, $
D_{\rm s}N_{\rm s} = D_{\rm i}N_{\rm i} = 0.040\pm 0.001$. We note
that a distribution with the number $ M $ of modes considerably
lower than one is highly peaked around the value $ n=0 $, which is
a consequence of specific form of the noise occurring in the
detection process. The obtained parameters reveal that the
measured weak twin beam was composed of on average 8.8 photon
pairs and 0.07 (0.15) noise signal (idler) photons. Its joint
signal-idler photon-number distributions $ p_{\rm si}(n_{\rm
s},n_{\rm i}) $ obtained by the maximum-likelihood approach as
well as the calibration method, together with the experimental
joint signal-idler photocount histogram $ f_{\rm si}(c_{\rm
s},c_{\rm i}) $ [see Fig.~\ref{fig2}(a)], are plotted in
Figs.~\ref{fig2}(b) and \ref{fig2}(c), respectively. Thus, the
analyzed twin beam contains tight (quantum) correlations between
the signal and idler photon numbers on one side, on the other side
its marginal signal and idler photon-number distributions are
multi-thermal, i.e. very classical
\cite{Haderka2005a,Avenhaus2010}. We note that quantum properties
of such weak noisy twin beams in multi-mode Gaussian states have
been theoretically analyzed in Ref.~\cite{Arkhipov2015} and their
nonclassicality invariant describing the behavior of their
entanglement on a beam-splitter has been discussed in
Refs.~\cite{Arkhipov2016b,Arkhipov2016}.

On the other hand, the application of the maximum-likelihood
approach provides a joint signal-idler photon-number distribution
$ p_{\rm si}(n_{\rm s},n_{\rm i}) $ as a steady state of the
following iteration procedure \cite{Dempster1977,PerinaJr2012}:
\begin{eqnarray} 
 p_{\rm si}^{(l+1)}(n_{\rm s},n_{\rm i}) &=& p_{\rm si}^{(l)}(n_{\rm s},n_{\rm i})
  \nonumber \\
 & & \hspace{-15mm} \mbox{} \times
  \sum_{c_{\rm s},c_{\rm i}} \frac{ f_{\rm si}(c_{\rm s},c_{\rm i})
  T_{\rm s}(c_{\rm s},n_{\rm s}) T_{\rm i}(c_{\rm i},n_{\rm i}) }{
  \sum_{n'_s,n'_i} T_{\rm s}(c_{\rm s},n'_{\rm s})
   T_{\rm i}(c_{\rm i},n'_{\rm i})
   p_{\rm si}^{(l)}(n'_{\rm s},n'_{\rm i}) }, \nonumber \\
 & & \hspace{15mm} l=0,1,\ldots .
\label{101}
\end{eqnarray}
The uniform initial distribution $  p_{\rm si}^{(0)}(n_{\rm
s},n_{\rm i}) $ is assumed in the iteration procedure. Compared to
the joint photon-number distribution $ p_{\rm si} $ obtained in
the calibration method, the distribution $ p_{\rm si} $ revealed
by the iteration procedure in Eq.~(\ref{101}) is broader, as
documented in Fig.~\ref{fig2}(b). This reflects slightly weaker
correlations between the signal and idler photon numbers (weaker
pairing of photons), i.e. greater mean numbers of the noise signal
and noise idler photons. As shown below, this is manifested when
considering various entanglement criteria.

\begin{figure}  
 \centerline{(a) \includegraphics[width=0.4\textwidth]{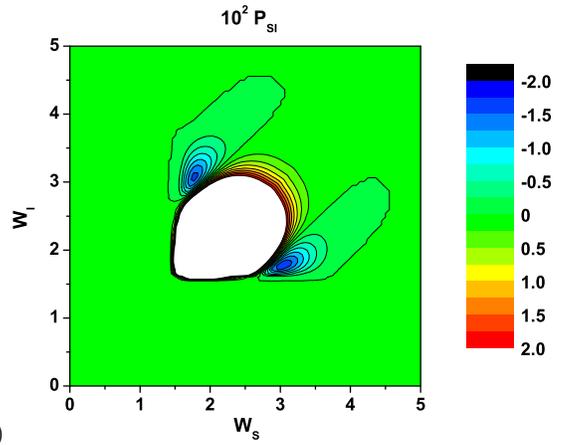}}
 \vspace{5mm}
 \centerline{(b) \includegraphics[width=0.4\textwidth]{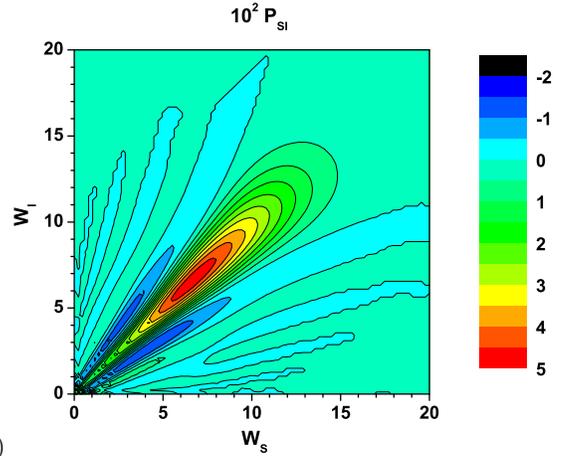}}
 \vspace{5mm}
 \centerline{(c) \includegraphics[width=0.4\textwidth]{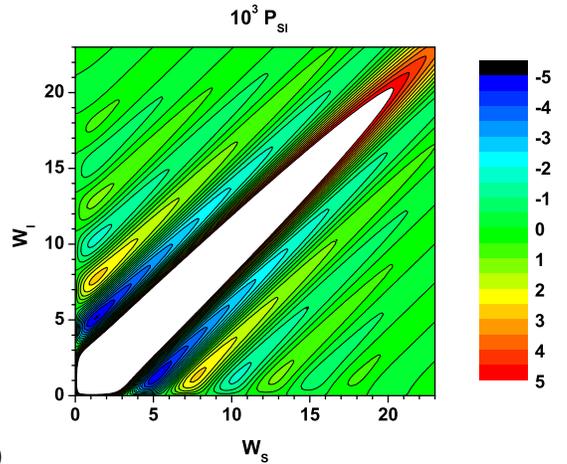}}
 \caption{Topo graphs of regularized quasi-distributions $ P_{\rm si}(W_{\rm s},W_{\rm i}) $
  of integrated intensities derived from (a) experimental photocount histogram $  f_{\rm si} $
  (via its multi-mode Gaussian fit) for the ordering parameter $ s = 1 $, (b)
  photon-number distribution $ p_{\rm si} $ reconstructed by the
  expectation-maximization approach (via the decomposition into
  the Laguerre polynomials) for $ s=0 $ and (c) photon-number distribution $ p_{\rm si} $
  reconstructed by the calibration method (via its multi-mode Gaussian fit) for $ s=0.1 $. In (a) [(c)], the maximum of
  $ P_{\rm si} $ inside the while area equals $ 3.6\times 10^{-2} $ [2.7]. When determining $ P_{\rm si}
  $ in (a) and (c), one effective mode comprising the whole signal (idler) beam has been assumed \cite{Haderka2005a,PerinaJr2013a,Perina1991,Perina2005}.}
\label{fig3}
\end{figure}

Nonclassicality (originating in local nonclassicality or
entanglement) of a bipartite field is inscribed into its joint
signal-idler quasi-distribution $ P_{\rm si}(W_{\rm s},W_{\rm i})
$ of integrated intensities $ W_{\rm s} $ and $ W_{\rm i} $ that
either attains negative values or even does not exist as a regular
analytical function \cite{Glauber1963,Sudarshan1963}. In our case,
we can obtain regularized forms of such quasi-distribution either
by direct evaluation (for a multi-mode Gaussian field)
\cite{Perina2005} or by using the decomposition of the
quasi-distribution into specific series of the Laguerre
polynomials with the weights derived from the appropriate joint
photocount and photon-number distributions \cite{Haderka2005a}. In
both cases, regularization of the quasi-distribution is provided
by the experimental noise. Parallel strips with negative values
are characteristic for the obtained regularized
quasi-distributions $ P_{\rm si}(W_{\rm s},W_{\rm i}) $ that are
plotted in Fig.~\ref{fig3}.

As the experimentally investigated noisy twin beams are mainly
composed of photon pairs and exhibit multi-mode thermal
photon-number statistics both in the signal and idler fields, they
cannot be locally nonclassical, but they exhibit the entanglement.
For this reason, we apply to the experimental histogram only the
GNCCa derived in the previous two sections. We analyze both the
joint experimental photocount histogram and the reconstructed
joint photon-number distributions arising in the calibration and
maximum-likelihood methods. We first pay attention to the GNCCa
containing intensity moments. To allow for certain comparison
among different GNCCa, we rewrite them in dimensionless units by
introducing the normalized GNCCa (denoted by tildes). They are
determined from the above written GNCCa by dividing them by
appropriate powers of the mean intensity $ \langle W\rangle =
(\langle W_{\rm s}\rangle + \langle W_{\rm i}\rangle )/2 $.
However, fair comparison of the performance of various GNCCa
containing intensity moments of different orders is based on the
corresponding (global) NCDs $ \tau $ introduced in Eq.~(\ref{84}).
In the second step and for comparison, we analyze the GNCCa given
in Eqs.~(\ref{93})---(\ref{98}) that use photon-number moments and
also some GNCCa involving the elements of photocount and
photon-number distributions.

\begin{figure}  
 \centerline{(a) \includegraphics[width=0.4\textwidth]{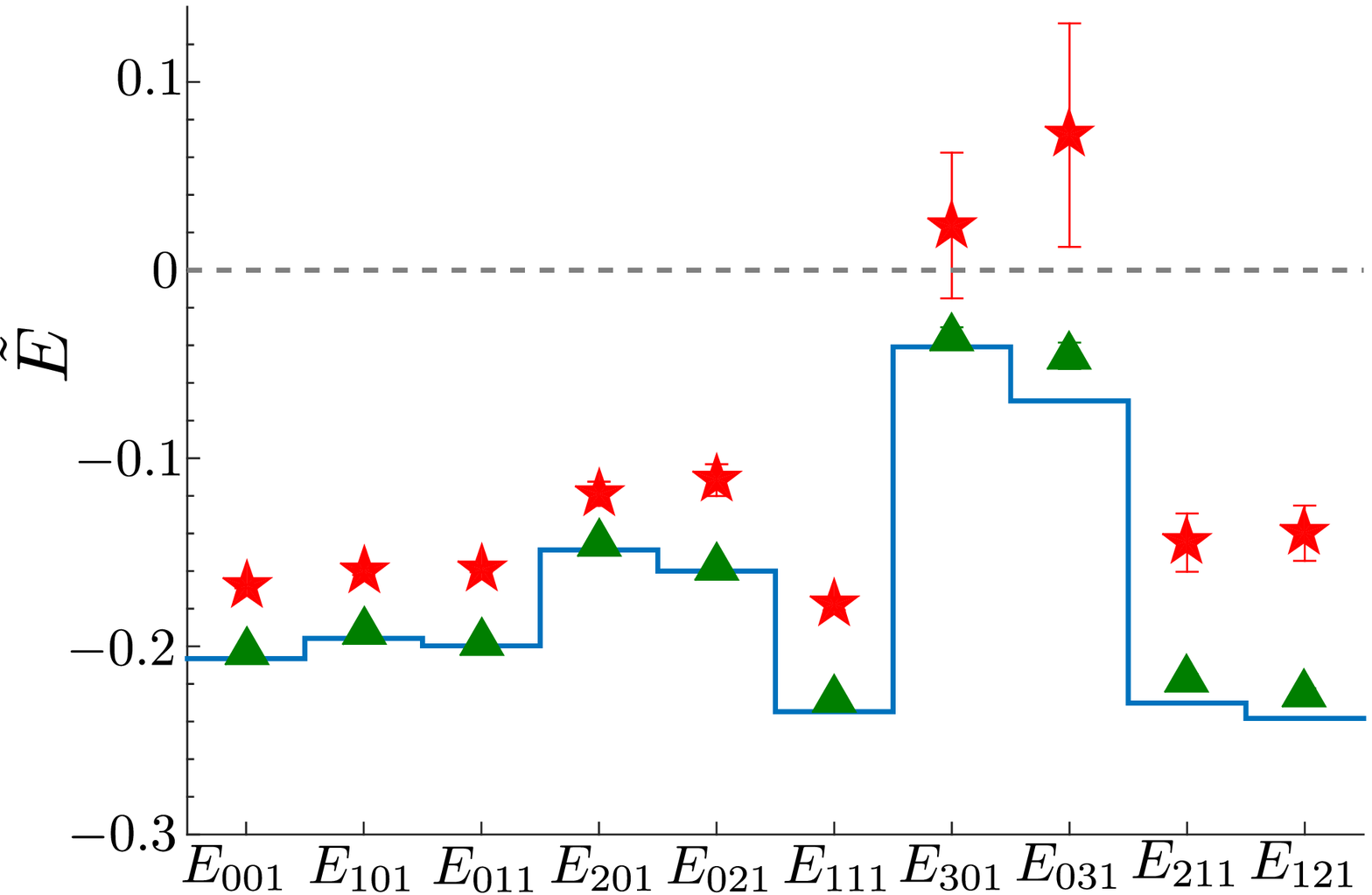}}
 \vspace{5mm}
 \centerline{(b) \includegraphics[width=0.4\textwidth]{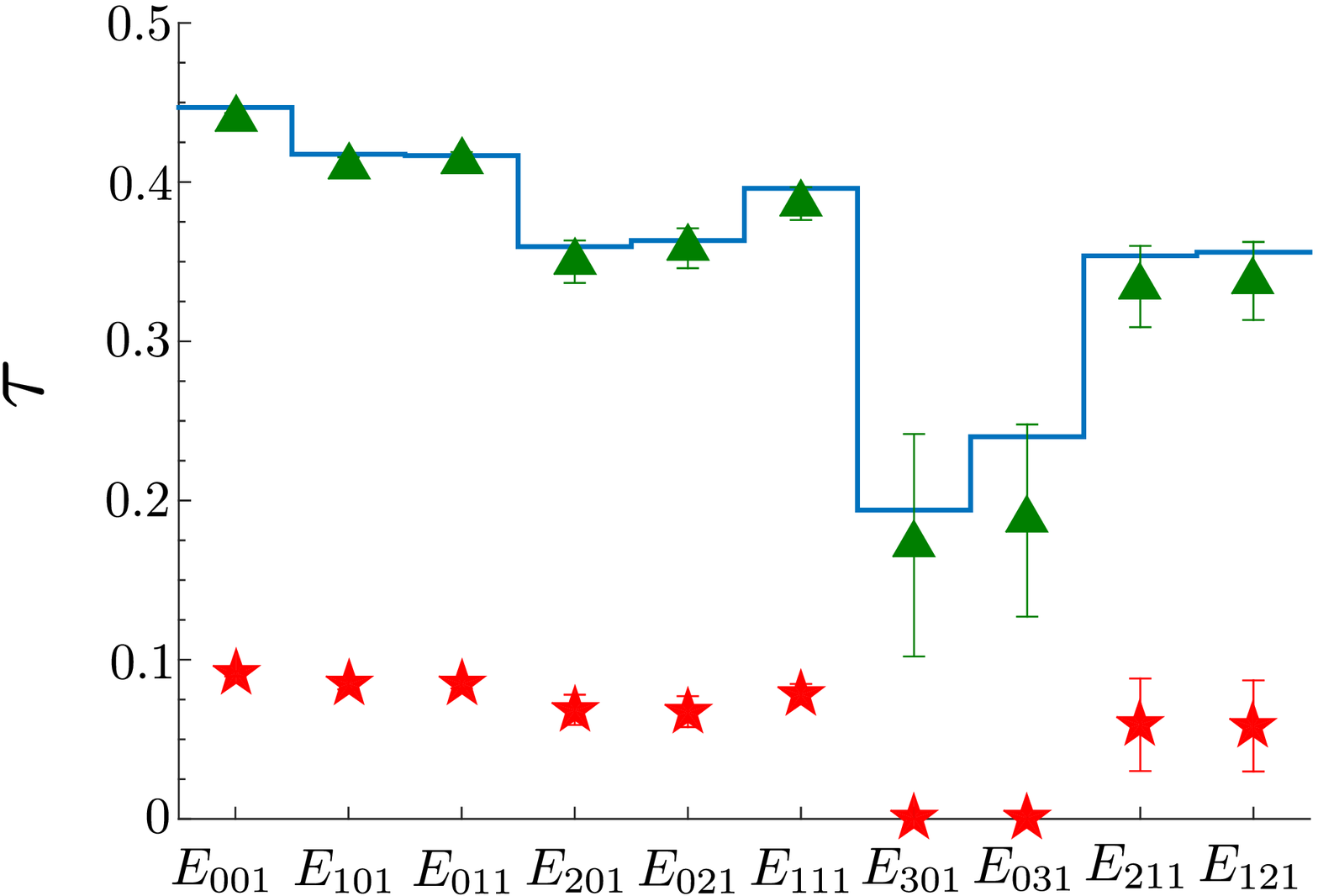}}
 \caption{(a) Normalized global nonclassicality criteria $ \tilde{E} $ defined in Eqs~(\ref{20})---(\ref{29})
  and (b) the corresponding nonclassicality depths $ \tau $. Values determined from the
  experimental photocount histogram are plotted by red asterisks whereas those appropriate for
  the reconstructed photon-number distribution reached by the
  maximum-likelihood (calibration) method are drawn by green triangles (blue solid curve).
  Some error bars are smaller than the used symbols.}
\label{fig4}
\end{figure}

In our opinion, the GNCCa $ E_{001}, \ldots, E_{121} $ given in
Eqs.~(\ref{20})---(\ref{29}) represent the basic set of GNCCa
suggested for the analysis of entanglement with the restriction up
to the fifth-order intensity moments. This is so because of their
simple forms and systematic inclusion of intensity moments of
different orders. Moreover, they can be derived in parallel from
the majorization theory and the inversion of simple classical
inequalities valid for non-negative polynomials. Also, the
simplest GNCC written in Eq.~(\ref{3}) was experimentally measured
already in 1991 \cite{Zou1991a}. The values of these GNCCa
determined for the experimental photocount histogram (red
asterisks), reconstructed photon-number distribution using the
maximum-likelihood method (green triangles) and reconstructed
photon-number distribution obtained by the calibration method
(blue solid curve) are plotted in Fig.~(\ref{fig4}), together with
the corresponding NCDs. Except for the GNCCa $ E_{301} $ and $
E_{031} $ applied to the photocount histogram, all other GNCCa
from this basic set are negative exhibiting the entanglement.
Positive values of the GNCCa $ E_{301} $ and $ E_{031} $ for the
photocount histogram are related to the occurrence of the
fifth-order marginal intensity moments in their definitions in
Eqs.~(\ref{26}) and (\ref{27}). Both types of the applied
reconstructions that partly remove the noise from the detected
photocount histogram lead to negative values of the GNCCa $
E_{301} $ and $ E_{031} $. The analysis of the corresponding NCDs
$ \tau $ reveals that the values of NCDs $ \tau $ decrease with
the increasing order of intensity moments involved in the GNCCa.
We note that similar decrease of the values of NCDs with the
increasing order of intensity moments has been observed in
\cite{PerinaJr2017} in case of LNCCa. Naturally, the values of
NCDs $ \tau $ are considerably greater for the reconstructed
photon-number distributions compared to the original experimental
photocount histogram.

\begin{figure}  
 \centerline{\includegraphics[width=0.4\textwidth]{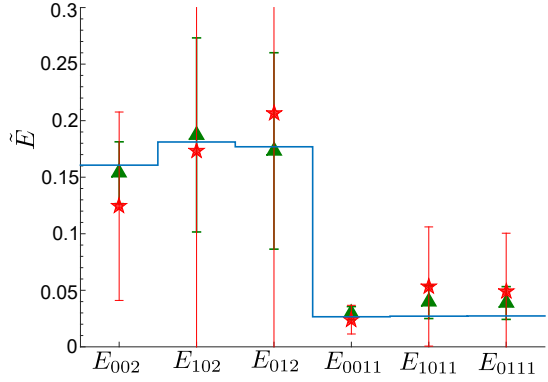}}
 \caption{Normalized global nonclassicality criteria $ \tilde{E} $ defined in
  Eqs~(\ref{30})---(\ref{32}), (\ref{65}), and (\ref{66}). For description,
  see the caption to Fig.~\ref{fig4}.}
\label{fig5}
\end{figure}

The basic set of GNCCa is accompanied by additional six GNCCa that
are derived similarly: $ E_{002} $ [Eq.~(\ref{30})], $ E_{102} $
[Eq.~(\ref{31})], $ E_{012} $ [Eq.~(\ref{32})], $ E_{0011} $
[Eq.~(\ref{66})], $ E_{1011} $ [Eq.~(\ref{65})], and $ E_{0111} $.
Unfortunately, none of these GNCCa indicates the entanglement in
the measured twin beam, as documented in Fig.~\ref{fig5}. Positive
values of the GNCCa $ E_{002} $, $ E_{102} $ and $ E_{012} $ can
again be related to the presence of the fourth- and fifth-order
marginal intensity moments in the definitions of these GNCCa. On
the other hand, the GNCCa $ E_{0011} $, $ E_{1011} $ and $
E_{0111} $ contain in their definitions the terms with two and
even three intensity moments in a product, which seriously limits
their ability to reveal entanglement.

Restricting our consideration to the fourth-order intensity
moments, the majorization theory provides five GNCCa [denoted by
symbol $ D $, Eqs.~(\ref{47})---(\ref{49})] for which products of
two intensity moments are characteristic, together with nine GNCCa
[denoted by symbol $ T $, Eqs.~(\ref{50})---(\ref{55})] containing
terms with up to three intensity moments in a product. All these
GNCCa indicate by their negative values the entanglement both in
the photocount histogram and the reconstructed photon-number
distributions, as documented in Figs.~\ref{fig6} and \ref{fig7}.
Mutual comparison of NCDs $ \tau $ for the GNCCa $ E $, $ D $ and
$ T $ plotted in turn in Figs.~{\ref{fig4}, {\ref{fig6} and
{\ref{fig7} reveals that the entanglement described by the GNCCa $
E $ is the most resistant against the noise, the GNCCa $ D $ are
considerably worse from this point of view and the resistance of
the GNCCa $ T $ against the noise is already weak. This behavior
can qualitatively be explained by the occurrence of multiple
products of intensity moments in the expressions giving the GNCCa
$ D $ and $ T $. These products do not naturally describe any
correlation and so their presence in the GNCCa only weakens the
ability of a given GNCCa to identify the entanglement.
\begin{figure}  
 \centerline{\includegraphics[width=0.22\textwidth]{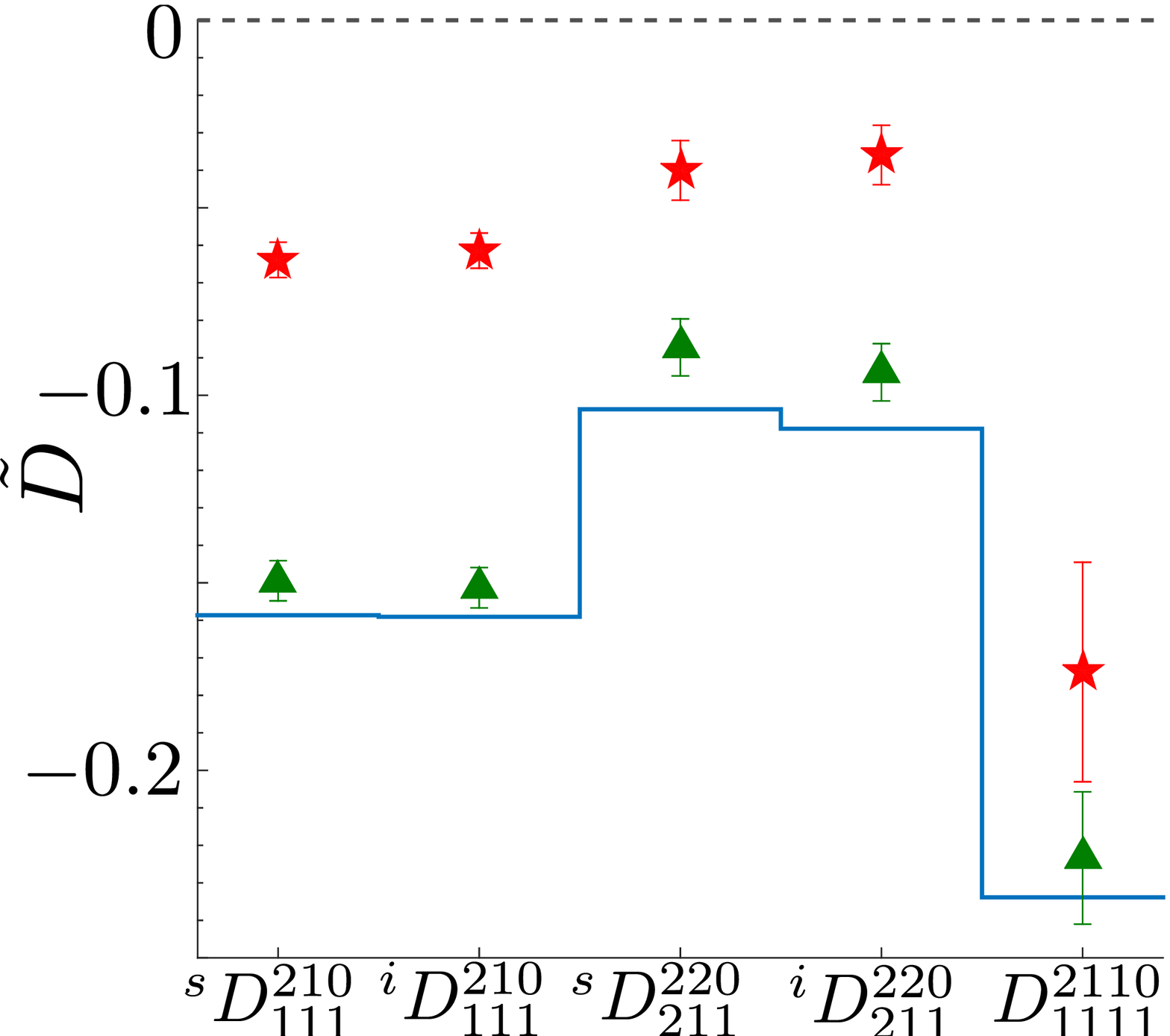}
 \hspace{2mm}
 \includegraphics[width=0.22\textwidth]{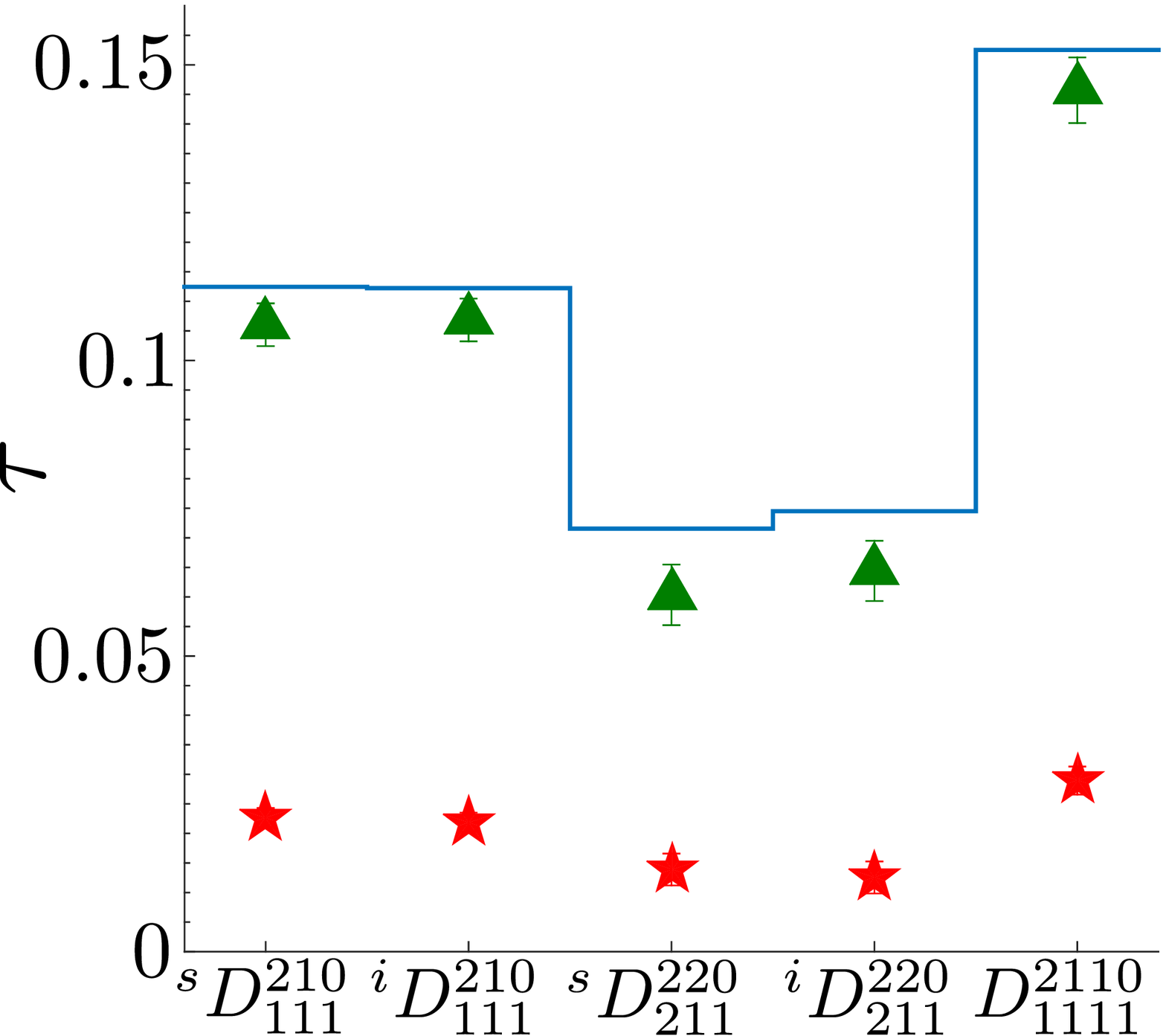}}
 \centerline{(a) \hspace{.25\textwidth} (b)}
 \caption{(a) Normalized global nonclassicality criteria $ \tilde{D} $ defined in Eqs~(\ref{47})---(\ref{49})
  and (b) the corresponding nonclassicality depths $ \tau $.
  For description, see the caption to Fig.~\ref{fig4}.}
\label{fig6}
\end{figure}
\begin{figure}  
 \centerline{(a) \includegraphics[width=0.4\textwidth]{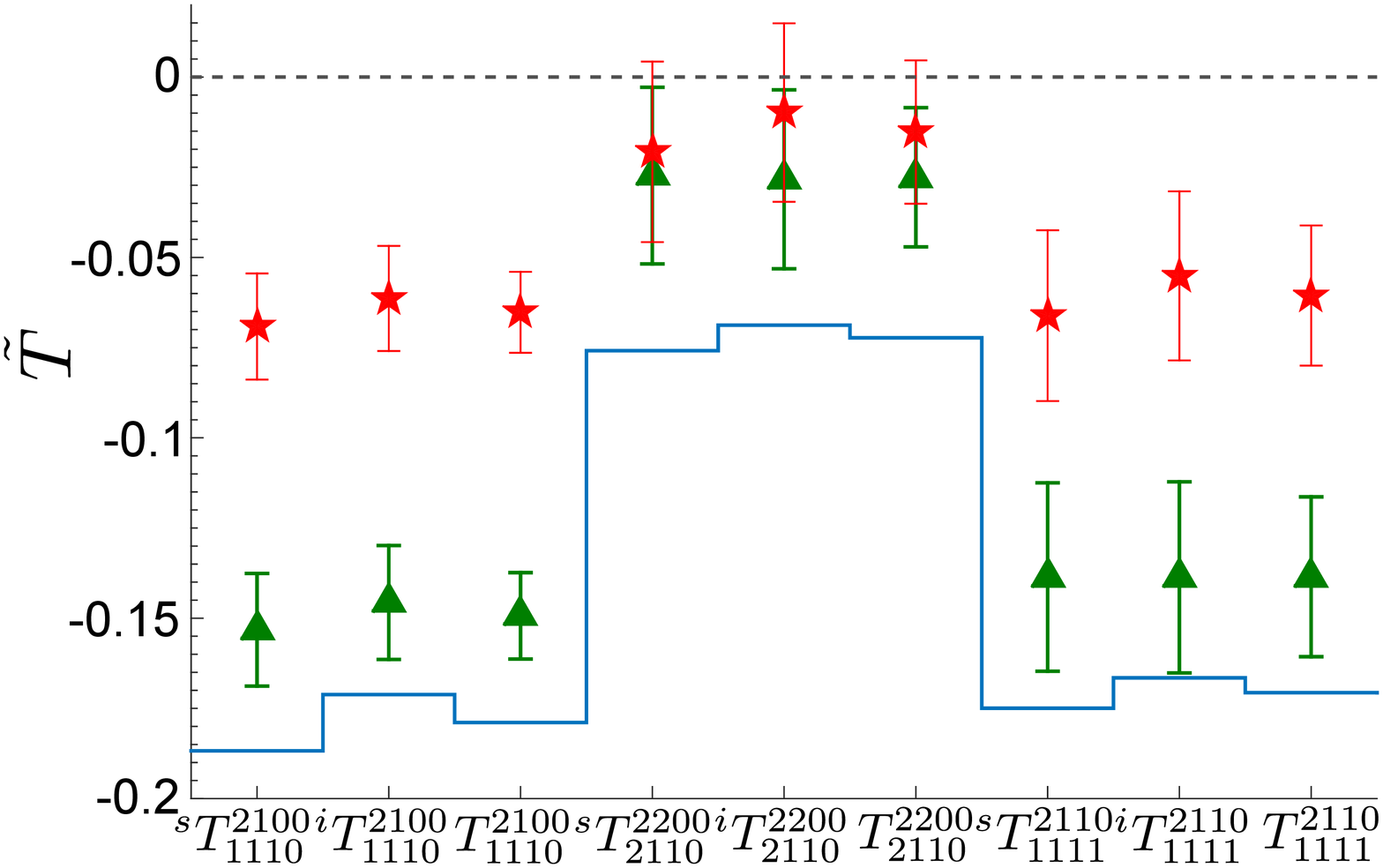}}
 \vspace{5mm}
 \centerline{(b) \includegraphics[width=0.4\textwidth]{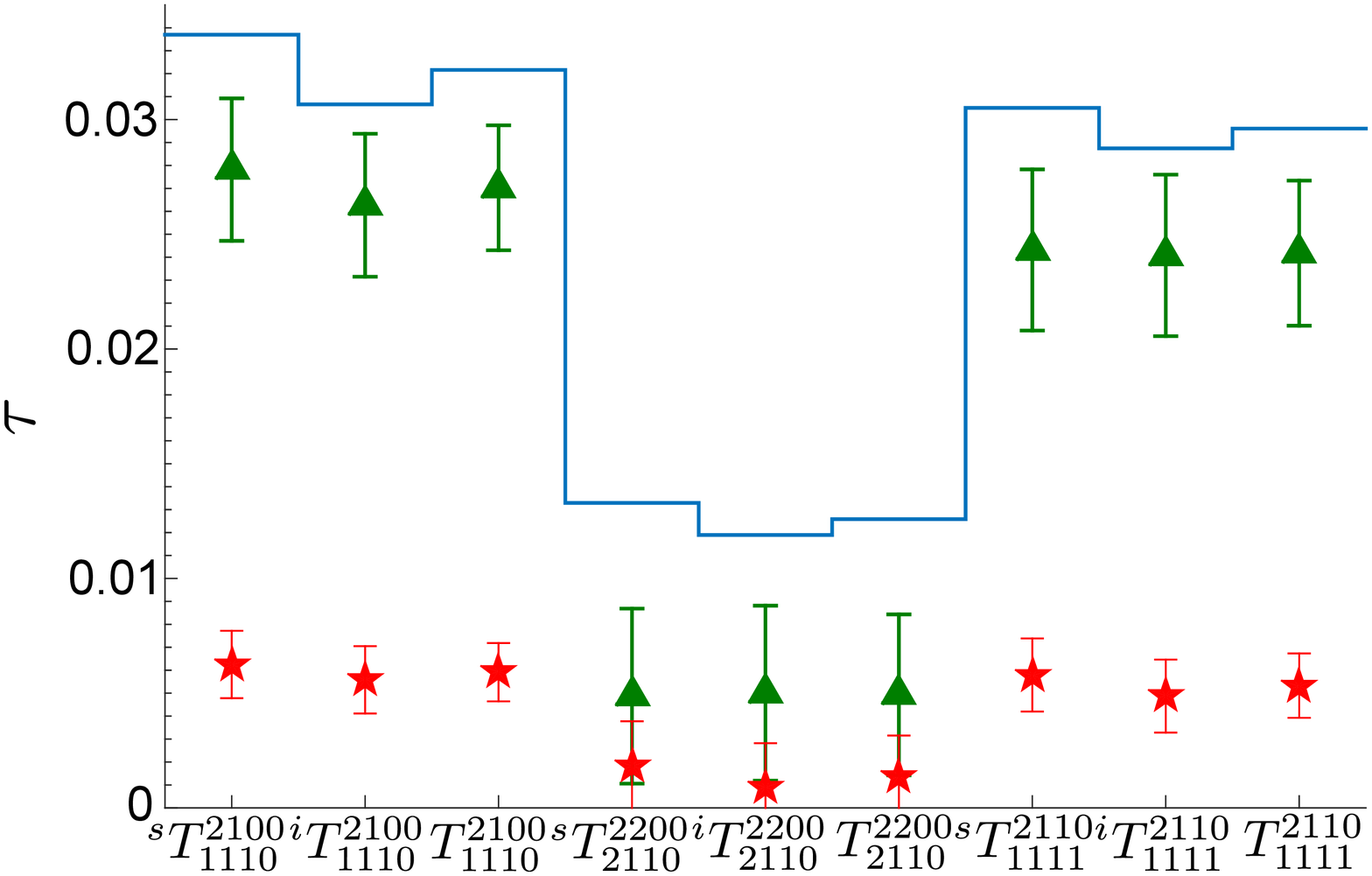}}
 \caption{(a) Normalized global nonclassicality criteria $ \tilde{T} $ defined in Eqs~(\ref{50})---(\ref{55})
  and (b) the corresponding nonclassicality depths $ \tau $.
  For description, see the caption to Fig.~\ref{fig4}.}
\label{fig7}
\end{figure}

The widely used matrix approach
\cite{Vogel2008,Miranowicz2010,Sperling2015} gives us three GNCCa
$ M_{1100} $ [Eq.~(\ref{73})], $ M_{1001} $ [Eq.~(\ref{74})] and $
M_{001001} $ [Eq.~(\ref{77})] for investigating entanglement,
provided that intensity moments up to the fifth order are taken
into account. For our experimental data, only the GNCCa $ M_{1001}
$ and $ M_{001001} $ identify entanglement (see Fig.~\ref{fig8}).
We note that negativity of the experimental GNCCa $ M_{1001} $ has
been reported in \cite{Allevi2012a}. The values of the
corresponding NCDs $ \tau $ plotted in Fig.~\ref{fig8} are
comparable to those characterizing the GNCCa $ E $ from the basic
set. This shows their high performance in identifying the
entanglement. A bit surprisingly the GNCC $ M_{1100} $ is
positive. In our opinion this is a consequence of the thermal
statistics of photon pairs. Loosely speaking and relying on the
quantum theory, we may define 'a photon-pair intensity' $ W_{\rm
si} \approx W_{\rm s}W_{\rm i} $ that allows us to rewrite
Eq.~(\ref{73}) in the form $ M_{1100} \approx \langle W_{\rm
si}^2\rangle - \langle W_{\rm si}\rangle^2 $ that explains
positivity of the GNCC $ M_{1100} $ for the analyzed weak twin
beam.

The Cauchy-Schwarz inequality provides two simple GNCCa not
mentioned above, $ C^{10}_{12} $ [Eq.~(\ref{80})] and $
C^{21}_{01} $ [Eq.~(\ref{82})] whose performance in revealing the
entanglement lies in between the GNCCa $ M_{1001} $ and $ M_{1100}
$ (see Fig.~\ref{fig8}). For the experimental twin beam, only the
GNCC $ C^{21}_{01} $ applied to the reconstructed photon-number
distributions indicates the entanglement. As the GNCC $
C^{10}_{12} $ is derived from the GNCC $ C^{21}_{01} $ by
substitution $ {\rm s} \leftrightarrow {\rm i} $, this
demonstrates strong sensitivity of both GNCCa to the level of
noise. The slightly lower mean of the signal noise photon number
compared to that of the idler field (0.07 versus 0.15) is
sufficient to observe the negative GNCC $ C^{21}_{01} $. For
comparison, we plot in Fig.~\ref{fig8} another two GNCCa $
D^{2200}_{1111} $ [Eq.~(\ref{75})] and $ D^{4000}_{1111} $
[Eq.~(\ref{76})] that also contain the cross-correlation intensity
moments $ \langle W_{\rm s}W_{\rm i}\rangle $ and $ \langle W_{\rm
s}^2W_{\rm i}^2\rangle $ and that are expressed as positive linear
combinations of the already analyzed GNCCa. However, their NCDs $
\tau $ are lower due to the additional terms with marginal
higher-order intensity moments occurring in their definitions
compared to the formulas for the GNCCa $ M $ written in
Eqs.~(\ref{74}) and (\ref{77}).
\begin{figure}  
 \centerline{(a) \includegraphics[width=0.4\textwidth]{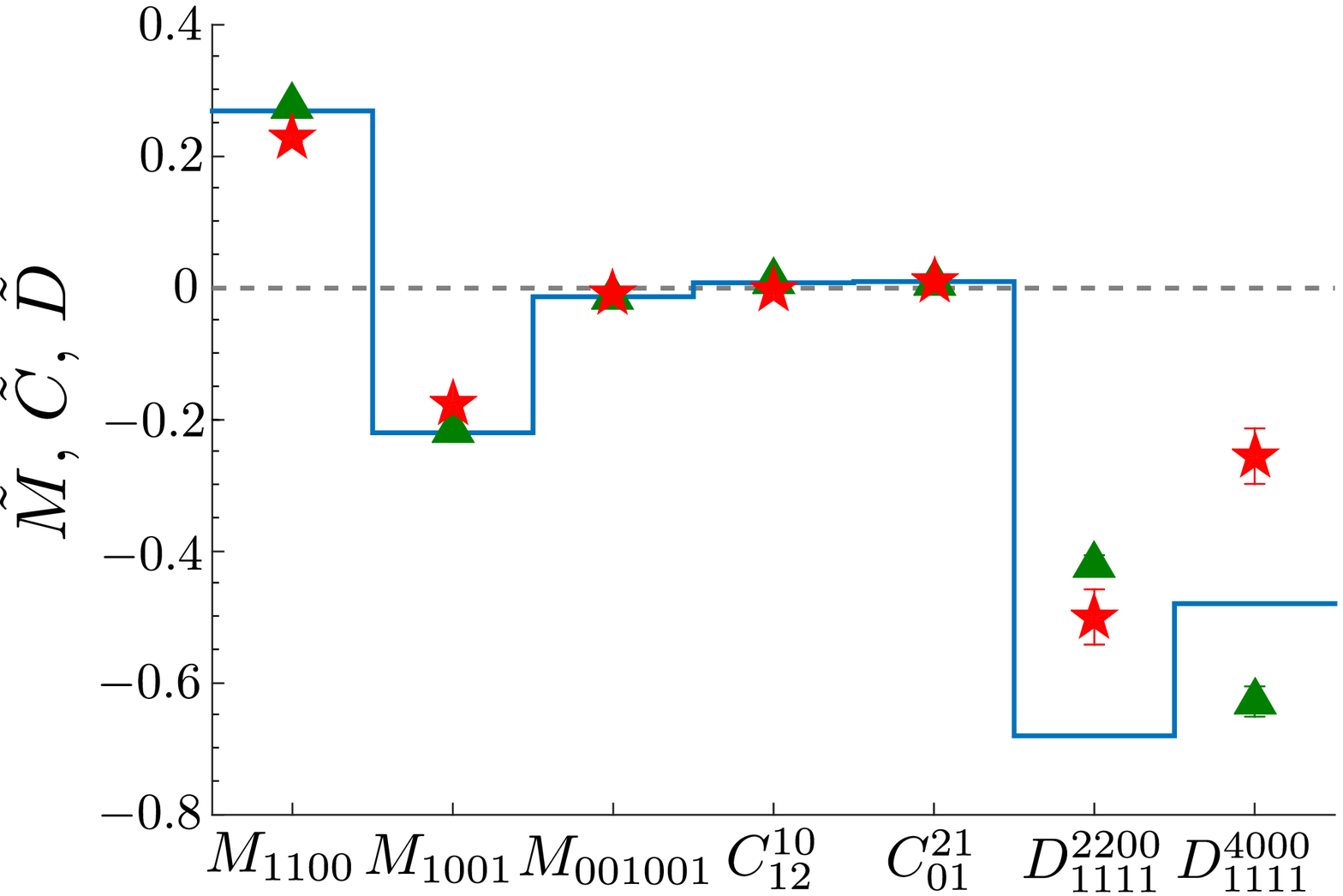}}
 \vspace{5mm}
 \centerline{(b) \includegraphics[width=0.4\textwidth]{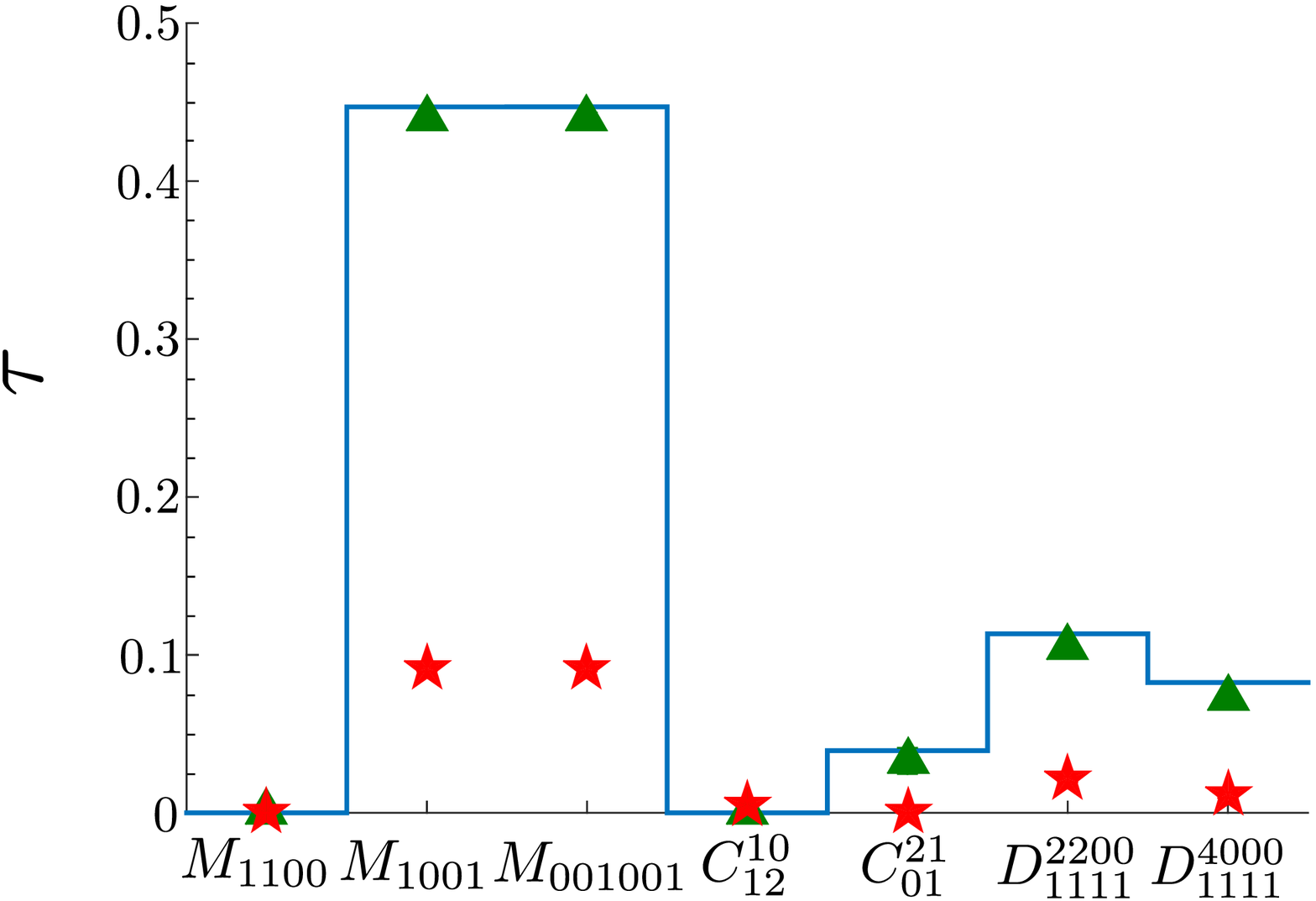}}
 \caption{(a) Normalized global nonclassicality criteria $ \tilde{M} $, $ \tilde{C} $ and $ \tilde{D} $
  defined in Eqs~(\ref{73})---(\ref{77}), (\ref{80}), and
  (\ref{82}) and (b) the corresponding nonclassicality depths $ \tau $.
  For description, see the caption to Fig.~\ref{fig4}.}
\label{fig8}
\end{figure}

All the above discussed GNCCa that are based on the intensity
moments can straightforwardly be converted into the corresponding
GNCCa that contain photocount and photon-number moments using the
linear relations between both types of moments quantified by the
Stirling numbers $ S $ [see Eq.~(\ref{92})]. This is more-or-less
formal for the reconstructed photon-number distributions. Contrary
to this, such GNCCa are useful and convenient when experimental
photocount histograms are analyzed. The reason is that these GNCCa
can directly be applied to the experimental data. This is why we
have suitably combined together various GNCCa written for the
intensity moments to arrive at a specific set of six simple GNCCa
$ N $ written in Eqs.~(\ref{93})---(\ref{98}). All of them have
been able to reveal the entanglement in the experimental
histogram, as documented in Fig.~\ref{fig9}. However, we note that
the GNCCa $ N $ are expressed as sums of intensity moments of
different orders and, as such, their structure is less transparent
compared to the original GNCCa based on the intensity moments.
\begin{figure}  
 \centerline{\includegraphics[width=0.37\textwidth]{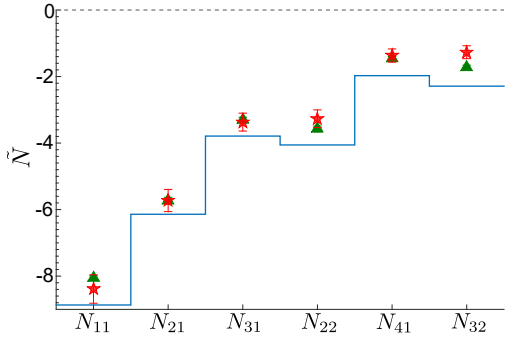}}
 \caption{Normalized global nonclassicality criteria $ \tilde{N} $ defined in Eqs~(\ref{93})---(\ref{98}).
  For description, see the caption to Fig.~\ref{fig4}.
  Normalization is done with respect to the corresponding
  quantities $ N^{\rm ref} $ determined for the factorized
  distribution $ P_{\rm s}(W_{\rm s})P_{\rm i}(W_{\rm i}) $.}
\label{fig9}
\end{figure}

The comparison of the results reached by the above discussed GNCCa
applied to the photon-number distributions reconstructed by the
maximum-likelihood approach and the calibration method reveals the
following. Negative values of the GNCCa, that reveal the
entanglement, reached by both approaches equal within the
experimental errors or the values provided by the
maximum-likelihood approach are greater than those reached by the
calibration method. In consequence, the corresponding NCDs from
both approaches coincide within the experimental errors or those
arising in the calibration method are greater. This behavior
naturally stems from the fact that the calibration method is more
efficient in removing the noise from the experimental data. This
is so as the calibration method works with a pre-defined form of
the photon-number distribution and applies it simultaneously to
the whole 2D experimental photocount histogram.

Finally, all the above written GNCCa as well as LNCCa can be
transformed into the corresponding GNCCa and LNCCa that involve
the elements of photocount histogram or reconstructed
photon-number distributions using the formal substitution written
in Eq.~(\ref{89}). The use of such GNCCa, however, needs different
approach compared to that applied to the GNCCa containing
intensity moments. Whereas only the intensity moments up to
certain order are useful owing to the increasing experimental
error with the increasing order of intensity moment, useful and
reliable GNCCa in case of the distributions involve their elements
(probabilities) having the highest available values. As both the
joint photocount histogram $ f_{\rm si} $ and the joint
reconstructed photon-number distributions $ p_{\rm si} $ have such
elements around the diagonal (see Fig.~\ref{fig2}), we consider
the GNCCa involving the elements at the diagonal
\cite{Lee1998,Waks2006} and the closest neighbor parallel lines,
as described in turn by functions $ F_{kk1} $, $ F_{(k+j)k1} $ and
$ F_{k(k+j)1} $, $ j=1,2 $, with the varying index $ k $ (see
Fig.~\ref{fig10}). The GNCCa $ F $ defined in Eq.~(\ref{90})
reveal reliably the entanglement via their negative values in the
area around the peaks of both the photocount histogram ($ k\approx
2 $) and reconstructed photon-number distributions ($ k \approx 9
$). We note that negative values of the GNCCa $ F_{(k+j)k1} $ and
$ F_{k(k+j)1} $ for $ j=2,\ldots $ [$ j=1,\ldots $] have not been
observed for the photon-number distribution reconstructed by the
maximum-likelihood [calibration] method which is a consequence of
its narrow 'cigar' shape clearly visible in Fig.~\ref{fig2}(b)
[\ref{fig2}(c)].
\begin{figure}  
 \centerline{(a) \includegraphics[width=0.3\textwidth]{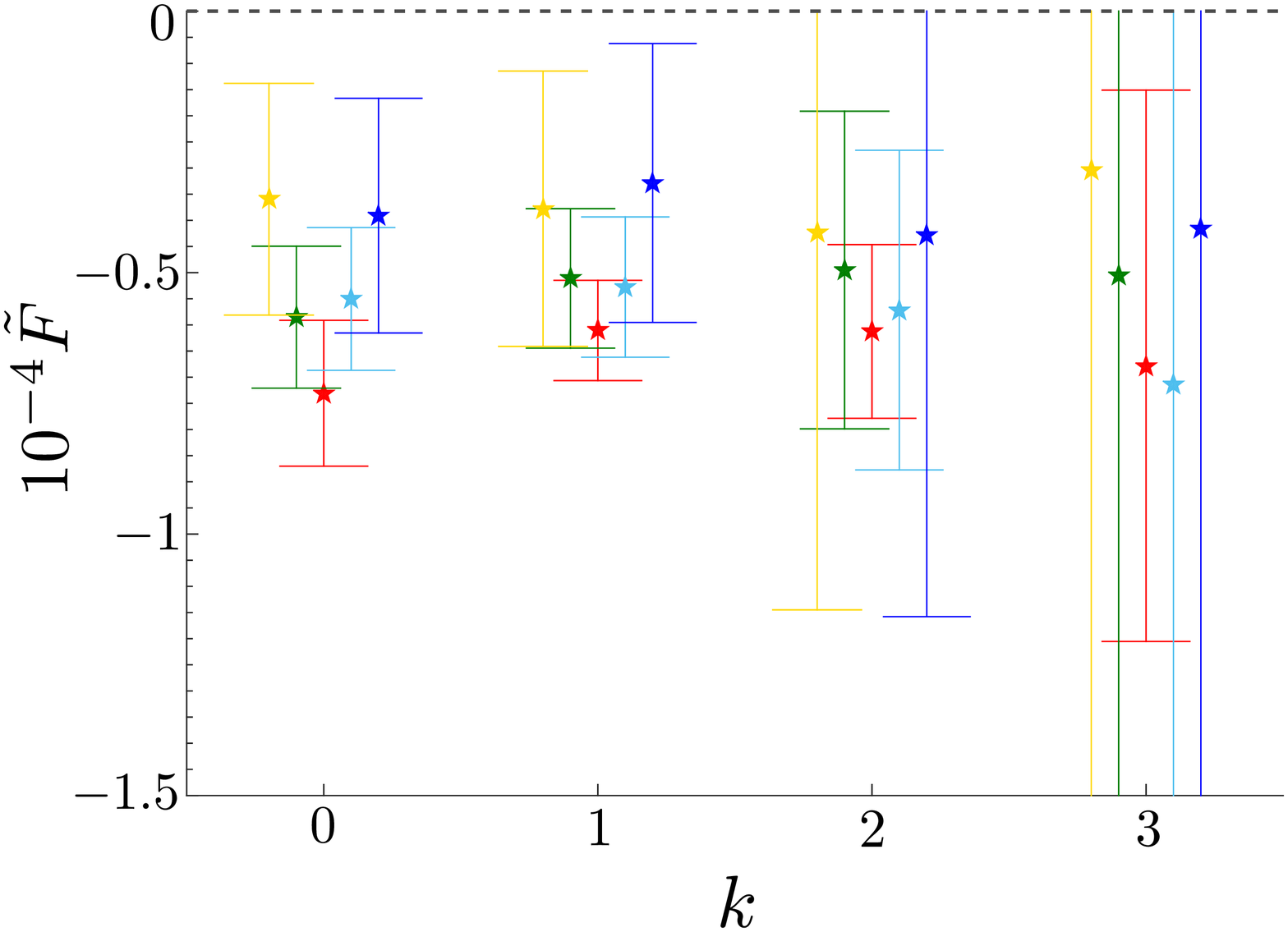}}
 \vspace{5mm}
 \centerline{(b) \includegraphics[width=0.3\textwidth]{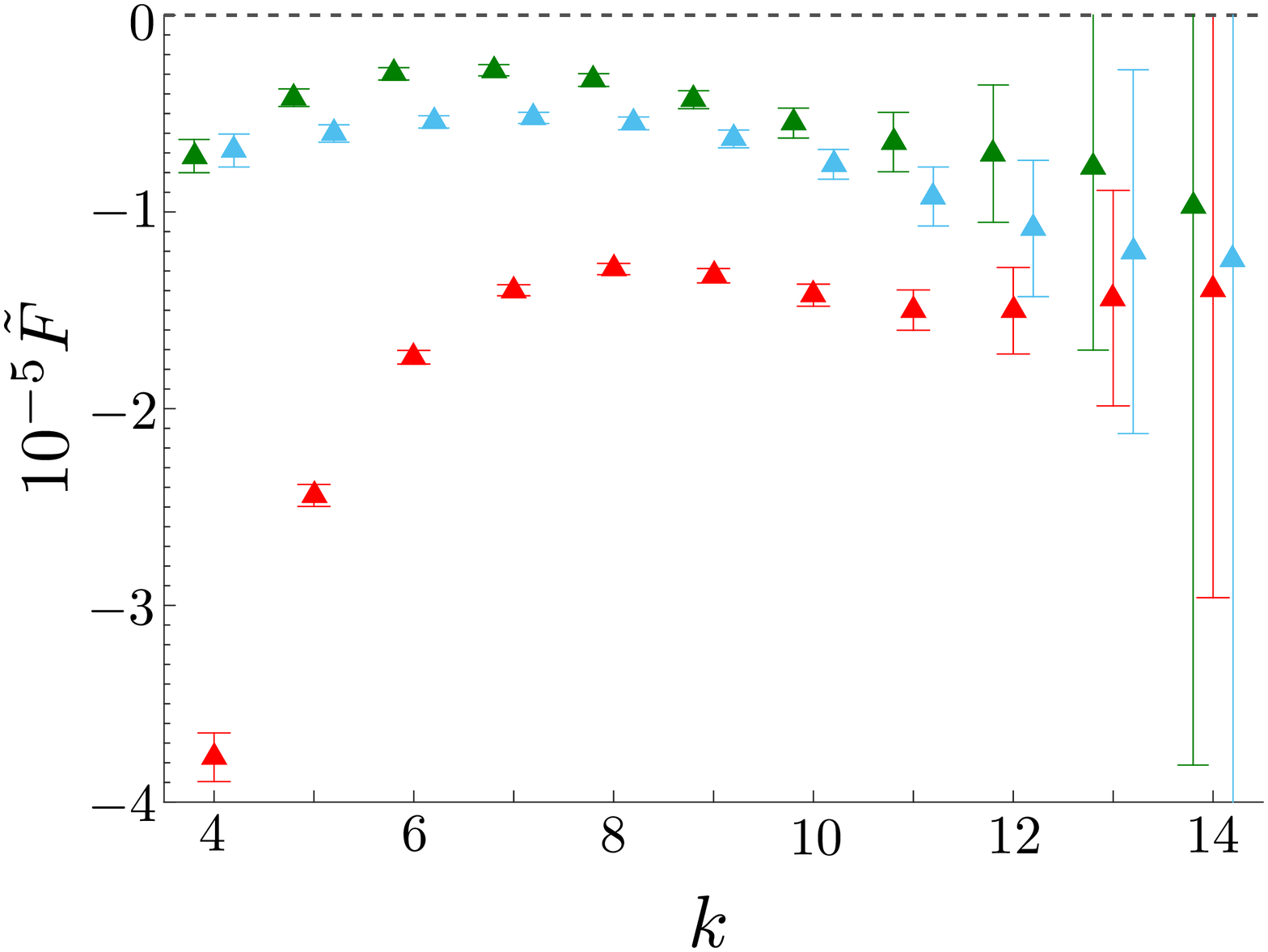}}
 \vspace{5mm}
 \centerline{(c) \includegraphics[width=0.3\textwidth]{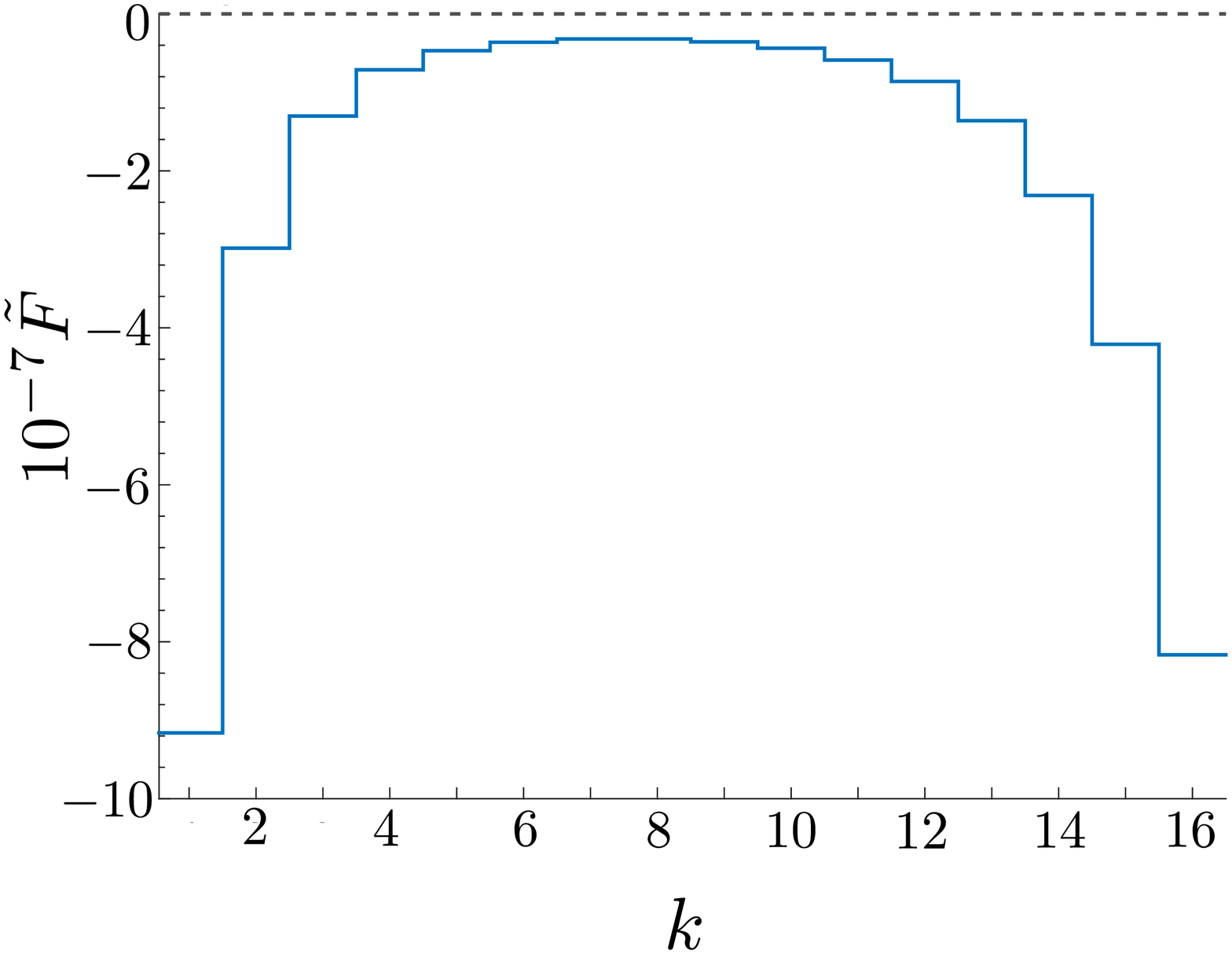}}
 \caption{Normalized global nonclassicality criteria $ \tilde{F}_{k'l'1} $ given in
  Eq.~(\ref{90}) for (a) experimental photocount histogram $  f_{\rm si} $
  and $ k'l' = kk $ (red asterisks), $ (k+1)k $
  (green), $ (k+2)k $ (yellow), $ k(k+1) $ (light blue), $ k(k+2) $ (dark blue), (b)
  photon-number distribution $ p_{\rm si} $ reconstructed by the
  maximum-likelihood approach and $ k'l' = kk $ (red triangles), $ (k+1)k $
  (green), $ k(k+1) $ (blue) and (c) photon-number distribution $ p_{\rm si} $
  reconstructed by the calibration method for $ k'l' = kk $ (solid blue
  curve);
  $ \tilde{F}_{kl1} \equiv [(k+1)(k+2)p_{\rm si}(k+2,l) + (l+1)(l+2)p_{\rm si}(k,l+2)
  - 2(k+1)(l+1)p_{\rm si}(k+1,l+1)] /[(k+1)(k+2)p_{\rm s}^{\rm P}(k+2)p_{\rm i}^{\rm P}(l)
   + (l+1)(l+2)p_{\rm s}^{\rm P}(k)p_{\rm i}^{\rm P}(l+2) - 2(k+1)(l+1)p_{\rm s}^{\rm P}(k+1)
   p_{\rm i}^{\rm P}(l+1)] $ and $ p_{a}^{\rm P}(n) $ is the
   Poissonian distribution with mean $ \langle n_a\rangle $
   normalized such that $ p_{a}^{\rm P}(0) = 1 $, $ a={\rm s,i} $.}
\label{fig10}
\end{figure}

\section{Conclusions}

We have derived numerous inequalities among the moments of
integrated intensities aimed at identifying local as well as
global nonclassicality using a) the majorization theory, b)
non-negative polynomials, c) the matrix approach based on
quadratic forms and d) the Cauchy-Schwarz inequality. We have
mutually compared different approaches, grouped the obtained
nonclassicality criteria according to their structure and tested
their performance on the experimental data characterizing a weak
twin beam with about nine photon pairs per pulse and small amount
of an additional noise. We have identified a basic set of ten
global nonclassicality criteria, that have revealed the
entanglement in the analyzed twin beam. We have also paid
attention to the counterparts of nonclassicality criteria written
in the moments of photocounts and photon numbers and also the
elements of photocount and photon-number distributions. We have
demonstrated their performance on the same experimental data. For
twin beams with low amount of the noise all three different kinds
of nonclassicality criteria represent a strong tool for revealing
the entanglement.

\acknowledgments The authors thank M. Hamar for his help with the
experiment. The authors were supported by GA \v{C}R (Project
No.~15-08971S) and M\v{S}MT \v{C}R (Project No.~LO1305). I.A. was
supported by GA \v{C}R (Project No.~17-23005Y) and UP (Project
No.~IGA\_PrF\_2017\_005).

\appendix

\section{Additional (redundant) nonclassicality criteria}

In Appendix~A, we summarize the nonclassicality criteria derived
from the majorization theory with polynomials written in three and
four variables and being redundant with respect to those presented
in the main text. This means that such LNCCa and GNCCa are
expressed as positive linear combinations of the LNCCa and GNCCa
written in the main text.

First, we summarize the redundant (and properly normalized) LNCCa
that complement the LNCCa contained in
Eqs.~(\ref{35})---(\ref{40}) and ({\ref{56})---(\ref{59}) ($
a={\rm s,i} $):
\begin{eqnarray}  
 ^{a}B^{200}_{110} &=& ^{a}L^{20}_{11} + B^{20}_{11} < 0,
  \label{A1} \\
 ^{a}B^{300}_{210} &=& ^{a}L^{30}_{21} + B^{30}_{21} < 0, \\
 ^{a}B^{400}_{310} &=& ^{a}L^{40}_{31} + B^{40}_{31} < 0, \\
 ^{a}B^{310}_{220} &=& ^{a}L^{31}_{22} + B^{31}_{22} < 0, \\
 ^{a}B^{2000}_{1100} &=& 2\; ^{a}L^{20}_{11} + B^{20}_{11} < 0, \\
 B^{2000}_{1100} &=& ^{\rm s}L^{20}_{11} + ^{\rm i}L^{20}_{11} + 2B^{20}_{11} < 0, \\
 ^{a}B^{3000}_{2100} &=& 2^{a}L^{30}_{21} + B^{30}_{21} < 0, \\
 B^{3000}_{2100} &=& ^{\rm s}L^{30}_{21} + ^{\rm i}L^{30}_{21} + 2B^{30}_{21} < 0, \\
 ^{a}B^{2100}_{1110} &=& \langle W_a\rangle \;^{a}L^{20}_{11} + ^{a}B^{210}_{111} < 0, \\
 B^{2100}_{1110} &=& ^{\rm s}B^{210}_{111} + ^{\rm i}B^{210}_{111} < 0, \\
 ^{a}B^{4000}_{3100} &=& 2\;^{a}L^{40}_{31} + B^{40}_{31} < 0, \\
 B^{4000}_{3100} &=& ^{\rm s}L^{40}_{31} + ^{\rm i}L^{40}_{31} + 2B^{40}_{31} < 0, \\
 ^{a}B^{3100}_{2200} &=& 2\;^{a}L^{31}_{22} + B^{31}_{22} < 0, \\
 B^{3100}_{2200} &=& ^{\rm s}L^{31}_{22} + ^{\rm i}L^{31}_{22} + 2B^{31}_{22} < 0, \\
 ^{a}B^{2200}_{2110} &=& \langle W_a^2\rangle \; ^{a}L^{20}_{11} + ^{a}B^{220}_{211} < 0, \\
 B^{2200}_{2110} &=& ^{\rm s}B^{220}_{211} + ^{\rm i}B^{220}_{211} < 0.
  \label{A16}
\end{eqnarray}

The redundant (and properly normalized) GNCCa containing the terms
with up to two intensity moments in a product attain the form ($
a={\rm s,i} $):
\begin{eqnarray}  
 ^{a}D^{200}_{110} &=& ^{a}L^{20}_{11} + (E_{001}+B^{20}_{11})/2 < 0 ,
   \label{A17} \\
 ^{a}D^{300}_{210} &=& 2\;^{a}L^{30}_{21} + E_{101} + E_{011} + B^{30}_{21} < 0 , \\
 ^{a}D^{400}_{310} &=& 2\;^{a}L^{40}_{31} + E_{201} + E_{111} + E_{021} + B^{40}_{31} < 0 , \nonumber \\
  & &  \\
 ^{a}D^{310}_{220} &=& 2\;^{a}L^{31}_{22} + E_{111} + B^{31}_{22} < 0 ,
  \label{A20} \\
 D^{2000}_{1100} &=& ^{\rm s}L^{20}_{11} + ^{\rm i}L^{20}_{11} + E_{001} + B^{20}_{11} < 0 ,
   \label{A21} \\
 D^{3000}_{2100} &=& ^{\rm s}L^{30}_{21} + ^{\rm i}L^{30}_{21} + E_{101} + E_{011} + B^{30}_{21} < 0 , \\
 D^{2100}_{1110} &=& (^{\rm s}D^{210}_{111} + ^{\rm i}D^{210}_{111})/2 < 0 , \\
 D^{4000}_{3100} &=& ^{\rm s}L^{40}_{31} + ^{\rm i}L^{40}_{31} + E_{201} + E_{111} + E_{021} \nonumber \\
  & & \mbox{} + B^{40}_{31} < 0 , \\
 D^{3100}_{2200} &=& ^{\rm s}L^{31}_{22} + ^{\rm i}L^{31}_{22} + E_{111} + B^{31}_{22} < 0 , \\
 D^{2200}_{2110} &=& ^{\rm s}D^{220}_{211} + ^{\rm i}D^{220}_{211} < 0.
  \label{A26}
\end{eqnarray}

Finally, the redundant (and properly normalized) GNCCa expressed
via triple products of intensity moments are derived as follows ($
a={\rm s,i} $):
\begin{eqnarray}  
 ^{a}T^{2000}_{1100} &=& 6\;^{a}L^{20}_{11} + E_{001}+ 2B^{20}_{11} < 0 ,
   \label{A27} \\
 T^{2000}_{1100} &=& ^{\rm s}L^{20}_{11} + ^{\rm i}L^{20}_{11} + (E_{001}+ 3B^{20}_{11})/2 < 0 , \\
 ^{a}T^{3000}_{2100} &=& 6\;^{a}L^{30}_{21} + E_{101} + E_{011} + 2B^{30}_{21} < 0 , \\
 T^{3000}_{2100} &=& 2\;^{\rm s}L^{30}_{21} + 2\;^{\rm i}L^{30}_{21} + E_{101} + E_{011} + 3B^{30}_{21} < 0 ,  \nonumber \\
  & & \\
 ^{a}T^{4000}_{3100} &=& 6\;^{a}L^{40}_{31} + E_{201} + E_{111} + E_{021} + 2B^{40}_{31} < 0 ,  \nonumber \\
  & & \\
 T^{4000}_{3100} &=& 2\;^{\rm s}L^{40}_{31} + 2\;^{\rm i}L^{40}_{31} + E_{201} + E_{111} + E_{021} \nonumber \\
  & & \mbox{} + 3B^{40}_{31} < 0 , \\
 ^{a}T^{3100}_{2200} &=& 6\;^{a}L^{31}_{22} + E_{111} + 2B^{31}_{22} < 0 , \\
 T^{3100}_{2200} &=& 2\;^{\rm s}L^{31}_{22} + 2\;^{\rm i}L^{31}_{22} + E_{111} + 3B^{31}_{22} < 0.
  \label{A34}
\end{eqnarray}


\end{document}